\documentclass{pas}
\usepackage[utf8]{inputenc}
\usepackage{multirow}
\usepackage{subfiles}
\usepackage{lipsum}
\usepackage{booktabs}
\usepackage{caption}
\usepackage{graphicx}
\usepackage{enumerate}
\usepackage{float}
\usepackage[dvipsnames]{xcolor}

\begin{document}

\lefttitle{Publications of the Astronomical Society of Australia}
\righttitle{D. J. Lyon \textit{et al.}}

\jnlPage{1}{4}
\jnlDoiYr{2021}
\doival{10.1017/pasa.xxxx.xx}

\articletitt{Research Paper}

\title{Decomposing Infrared Luminosity Functions into Star-Forming and AGN Components using CIGALE}

\author{\gn{Daniel J.} \sn{Lyon}$^{1}$,  
        \gn{Michael J.} \sn{Cowley}$^{1,2}$,
        \gn{Oliver} \sn{Pye}$^{1}$,
        and \gn{Andrew M.} \sn{Hopkins}$^{3}$}

\affil{
$^1$School of Chemistry and Physics, 2 George St, Queensland University of Technology, Brisbane, QLD 4000, Australia \\ 
$^2$Centre for Astrophysics, University of Southern Queensland, West Street, Toowoomba,
QLD 4350, Australia\\
$^3$School of Mathematical and Physical Sciences, 12 Wally’s Walk, Macquarie University, NSW 2109, Australia}

\corresp{Daniel Lyon, Email: daniellyon31@gmail.com}

\citeauth{Author1 C and Author2 C, an open-source python tool for simulations of source recovery and completeness in galaxy surveys. {\it Publications of the Astronomical Society of Australia} {\bf 00}, 1--12. https://doi.org/10.1017/pasa.xxxx.xx}

\history{(Received: 11 October 2024; revised xx xx xxxx; accepted xx xx xxxx)}

\begin{abstract}
    This study presents a comprehensive analysis of the infrared (IR) luminosity functions (LF) of star-forming (SF) galaxies and active galactic nuclei (AGN) using data from the ZFOURGE survey. We employ CIGALE to decompose the spectral energy distribution (SED) of galaxies into SF and AGN components to investigate the co-evolution of these processes at higher redshifts and fainter luminosities. Our CIGALE-derived SF and AGN LFs are generally consistent with previous studies, with an enhancement at the faint end of the AGN LFs. We attribute this to CIGALE's capability to recover low-luminosity AGN more accurately, which may be underrepresented in other works. We find evidence for a significant evolutionary epoch for AGN activity below $z \approx 2$, comparable to the peak of cosmic star formation at $z \approx 2$, which we also recover well. Based on our results, the gas supply in the early universe favoured the formation of brighter star-forming galaxies from high-redshift until $z=2$, below which the gas for SF becomes increasingly exhausted. In contrast, AGN activity peaked earlier and declined more gradually, suggesting a possible feedback scenario in which AGN positively influence SF. 
\end{abstract}


\maketitle


\section{Introduction} \label{Sec: Intro}
The distribution of galaxies and their luminosities have previously derived powerful constraints on galaxy evolution \citep{binggeli_luminosity_1988, benson_what_2003, rodighiero_mid-_2010, gruppioni_herschel_2013}. One of the direct ways of measuring the distribution of galaxies is with the luminosity function \citep{schechter_analytic_1976, saunders_60-mum_1990}. Luminosity functions (LFs) are statistical distributions that describe the spatial density of astronomical objects and are a fundamental tool for quantifying their evolution across cosmic time scales \citep{dai_mid-infrared_2009, han_evolution_2012, wylezalek_galaxy_2014}. The use of LFs in galaxy evolution studies has uncovered a wealth of information revealing the intricate processes governing star formation (SF), galaxy mergers, and the growth of supermassive black holes (SMBH, $M_{BH} > 10^{6}\ M_{\odot}$) across cosmic time \citep{caputi_infrared_2007, hopkins_observational_2007, magnelli_deepest_2013, delvecchio_tracing_2014, hernan-caballero_resolving_2015}. Specifically, many such studies find a strong correlation between the activity of the central SMBH and the star formation rate (SFR) \citep{hopkins_cosmological_2008, merloni_synthesis_2008}. 

Active galactic nuclei (AGN) are actively accreting SMBHs, whereby massive quantities of gas and dust power their growth \citep{hopkins_cosmological_2008, han_evolution_2012, toba_9_2013, brown_infrared_2019}. It is widely accepted that most galaxies, particularly those with significant bulges, host a SMBH at their centre \citep{gruppioni_modelling_2011, han_evolution_2012, brown_infrared_2019}. While not all central SMBHs are currently active, such as Sagittarius A$^{*}$ at the centre of our own Milky Way \citep{event_horizon_telescope_collaboration_first_2022}, most galaxies have likely experienced the influence of an AGN at some point in their history \citep{gruppioni_modelling_2011}. SF in galaxies similarly requires an extensive reservoir of cool gas to operate \citep{schawinski_observational_2007, cicone_massive_2014}. The same interstellar material that powers SF can also fuel AGN growth, leading to an inherent link between these two processes \citep{hopkins_cosmological_2008, brown_infrared_2019}. This connection has fuelled an ongoing debate over whether AGN activity enhances SF by triggering gas inflows or diminishes it through feedback mechanisms that deplete gas reservoirs \citep{grazian_galaxy_2015, fiore_agn_2017}.

Understanding the role of AGN in galaxy evolution is essential, as these processes regulate the growth of the SMBH and the host galaxy's development. Studies have shown that the most luminous AGN are often preceded by periods of intense SF \citep{kauffmann_host_2003, hopkins_cosmological_2008, hopkins_how_2010} suggesting a co-evolutionary relationship. Some of this activity is seen as AGN-driven outflow winds that can expel the interstellar medium (ISM) from the galaxy \citep{schawinski_observational_2007, cicone_massive_2014, fiore_agn_2017}, thereby starving the galaxy of the cold gas needed for both SF and AGN fueling, ultimately shutting down both processes \citep{hopkins_how_2010}. However, \cite{silk_unleashing_2013} suggested that AGN-driven winds may, under certain conditions, compress gas and dust, thereby enhancing SF. This scenario aligns with findings from \citet{cowley_zfourge_2016}, which indicate that AGN-dominated systems tend to have higher specific star formation rates, suggesting that SF and AGN activity can co-exist in certain environments.

Both SF and AGN activity releases an enormous amount of energy across the entire electromagnetic spectrum, from radio waves to gamma rays \citep{ho_spectral_1999, huang_local_2007, silva_modelling_2011, gruppioni_modelling_2011}. LFs at various wavelengths have been used to place powerful constraints on evolutionary models, as seen in \cite{aird_evolution_2015, alqasim_new_2023} (X-ray), \cite{yuan_determining_2018} (Radio), \cite{page_ultraviolet_2021} (UV), and \cite{cool_galaxy_2012} (optical). However, the properties of Infrared (IR) light make it the ideal regime for studying SF and AGN LFs as both processes are often dusty and obscured (optically thick) \citep{wu_mid-infrared_2011, han_evolution_2012}. Dust extinction absorbs the outgoing X-ray, optical, and UV wavelengths and re-emits the radiation in the IR domain \citep{fu_decomposing_2010, toba_9_2013, oconnor_luminosity_2016, symeonidis_agn_2021}. AGN activity is closely correlated with IR luminosity because the IR emission often traces dust heated by AGN \citep{kauffmann_host_2003, wu_mid-infrared_2011, symeonidis_what_2019, symeonidis_agn_2021}. However, this can create a bias against detecting faint AGN. This bias may also explain why AGN feedback appears to be more prevalent at lower redshifts \citep{katsianis_evolution_2017, pouliasis_obscured_2020}. The relationship between IR luminosity and SFR is complex \citep{symeonidis_agn_2021} and becomes weaker at brighter luminosities and higher redshifts, where AGN contribution to the IR emission increases \citep{wu_mid-infrared_2011}. 

Most studies focus on galaxy or SF LFs, which trace the evolution of galaxies \citep{tempel_tracing_2011, cool_galaxy_2012}, but often neglect the co-evolution with AGN \citep{fotopoulou_5-10_2016, symeonidis_agn_2021, finkelstein_coevolution_2022} even though galaxies are known to be significantly influenced by this relationship \citep{hopkins_cosmological_2008, fiore_agn_2017}. The population of obscured, dusty IR AGN and SFGs play a crucial role in constraining galaxy evolution models \citep{gruppioni_modelling_2011}. Analysing the SF and AGN LFs is paramount to understanding the complex processes driving galaxy evolution. SF LFs allow us to quantify the cosmic star formation rate history and the star-formation rate density. At the same time, the AGN LF provides insight into gas reservoirs not utilised by SF, particularly in the context of feedback mechanisms.

Since SF and AGN activity are so tightly coupled, distinguishing between the two components can be challenging. Traditional methods of selecting AGN, such as colour-colour diagnostics \citep{lacy_obscured_2004}, X-ray dominated \citep{szokoly_chandra_2004}, and radio-dominated \citep{rees_radio_2016} approaches, often overlook faint AGN due to biases inherent in their selection criteria \citep{thorne_deep_2022}. This limitation highlights the need for more refined methods of identifying AGN to better constrain galaxy evolution. In this paper, we use \texttt{Code Investigating Galaxy Emission} (CIGALE, \citealp{burgarella_star_2005, noll_analysis_2009, boquien_cigale_2019}) to decompose the IR Spectral Energy Distribution (SED) of ZFOURGE galaxies \citep{straatman_fourstar_2016} to generate and analyse both the SF and AGN LFs. Incorporating SED decomposition is crucial for interpreting the co-evolution of galaxies with AGN. SEDs characterise many ongoing processes such as the SF and AGN components, as well as dust attenuation and gas heating \citep{ho_spectral_1999, huang_local_2007, silva_modelling_2011, gruppioni_modelling_2011}. We focus on the SF and AGN components to probe galaxy evolution directly. SED decomposition will allow us to independently quantify SF and AGN evolution while minimising the bias against faint AGN.

To achieve this, we leverage data from the ZFOURGE survey, which probes galaxies at higher redshifts and fainter luminosities, allowing for more precise constraints on the evolution of the SFR and AGN activity across cosmic time. By decomposing the SED of each galaxy, we can disentangle the contributions of SF and AGN to galaxy evolution, providing a clearer picture of how these processes interact. Our approach minimises biases against faint AGN and improves the accuracy of the derived LFs. In section \ref{Sec: The ZFOURGE Survey}, we introduce the ZFOURGE Survey and an overview of our data generation and reduction. In section \ref{Sec: CIGALE}, we overview CIGALE and how we performed the SED decomposition process to disentangle SF and AGN luminosities. In section \ref{Sec: Luminosity Functions}, we provide a brief outline of how we calculate the LFs, the errors, and the functional fitting process. Finally, in section \ref{Sec: Discussion}, we discuss the results of our LFs, the parameter evolution, luminosity density, and luminosity class evolution in the broader context of galaxy co-evolution with AGN and SFR. Throughout this paper we adopt a cosmology of $H_0 = 70$ km $\mathrm{s^{-1}\ Mpc^{-1}}$, $\Omega_m=0.7$, and $\Omega_\Lambda=0.3$.

\section{The ZFOURGE Survey} \label{Sec: The ZFOURGE Survey}
\subsection{Overview}

This study utilises the 2017 release\footnote{Available for download at \href{https://zfourge.tamu.edu/}{zfourge.tamu.edu.}} of the ZFOURGE survey \citep{straatman_fourstar_2016}, which offers a unique combination of depth and wavelength coverage essential for probing high-redshift galaxies and constructing accurate LFs. ZFOURGE consists of approximately 70,000 galaxies at redshifts greater than 0.1, covering three major 11$\times$11 arcminute fields: the Chandra Deep Field South (CDFS) \citep{giacconi_chandra_2002}, the field observed by the Cosmic Evolution Survey (COSMOS) \citep{scoville_cosmic_2007}, and the CANDELS Ultra Deep Survey (UDS) \citep{lawrence_ukirt_2007}. These galaxies were observed using the near-infrared FourStar imager \citep{persson_fourstar_2013} mounted on the 6.5-m Magellan Baade Telescope at the Las Campanas Observatory in Chile. 

ZFOURGE employs deep near-infrared imaging with multiple medium-band filters (\textit{J}$_{1}$, \textit{J}$_2$, \textit{J}$_{3}$, \textit{H}$_{l}$, \textit{H}$_{s}$) and a broad-band \textit{K}$_{s}$ filter. The imaging spans 1.0 to 1.8 $\mu$m and achieves 5$\sigma$ point-source limiting depths of 26 AB mag in the \textit{J} medium-bands and 25 AB mag in the \textit{H} and \textit{K}$_{s}$ bands \citep{spitler_first_2012}. These filters yield well-constrained photometric redshifts, particularly effective for sources within the redshift range of 1 to 4 \citep{spitler_first_2012}. ZFOURGE data is supplemented by public data from HST/WFC3 F160W and F125W imaging from the CANDELS survey, Spitzer/Infrared Array Camera (IRAC), and Herschel/Photodetector Array Camera and Spectrometer (PACS). For a detailed description of the data and methodology, refer to \cite{straatman_fourstar_2016}.

\subsection{Sample Selection} \label{Sec: Sample Selection}
\subsubsection{ZFOURGE Sample} \label{Sec: Galaxy LF Selection}
ZFOURGE is a near-infrared-selected survey and, as such, may be biased against the most heavily dust-obscured galaxies, particularly at higher redshifts where the observed near-IR corresponds to rest-frame optical or ultraviolet wavelengths. This limitation is well established in the literature (e.g., \citealp{fu_decomposing_2010, grazian_galaxy_2015}) and is an inherent selection effect of deep near-IR surveys. As a result, the sample may underrepresent certain populations of extremely obscured star-forming galaxies and AGN. Nonetheless, the dataset includes a substantial number of dusty sources across all fields (e.g., \citealp{spitler_exploring_2014}), and many highly obscured systems, particularly those with high intrinsic luminosities, are still likely to be included due to their brightness in mid-IR or even near-IR bands (e.g., MIPS 24$\mu m$). 

To ensure the selection of high-quality galaxies and minimise errors in our analysis, we adopt the ZFOURGE quality flag \texttt{Use=1}, as defined by \cite{straatman_fourstar_2016}. This flag selects galaxies with reliable photometry and redshift measurements, resulting in a starting sample of 37,647 galaxies.

To ensure a physically meaningful sample, we excluded sources with negative infrared luminosities (LIR) derived from the \cite{wuyts_fireworks_2008} template fitting. These negative values arise from noise-dominated fluxes, particularly at 24–160$\mu m$, where photometric fluctuations result in unphysical luminosity estimates. Their removal introduces a minor level of incompleteness, but does not impact the reliability of the luminosity functions in the well-sampled, physically meaningful regime. After excluding galaxies with unphysical bolometric IR luminosities ($L_{IR} < 0$), the sample is reduced to 22,997 galaxies. 

We apply a redshift cut, restricting the sample to $0 \leq z \leq 6$ since only 28 galaxies exist at $z > 6$ (yielding  22,967 galaxies). This redshift range enables us to observe the evolution of galaxies during some of the most critical cosmic periods, specifically around $1 < z < 3$ \citep{gruppioni_modelling_2011, wylezalek_galaxy_2014} where galaxy luminosity density peaks \citep{assef_mid-ir-_2011}.

To ensure robustness, we calculate the estimated bolometric IR flux in each sample using the luminosity-distance equation and apply an 80\% completeness flux cut. Refer to section \ref{Sec: IR_Luminosity} for a description on how the bolometric IR luminosities were derived. This reduces the impact of noise and observational limits while preserving a large enough sample for LF construction. The final ZFOURGE sample includes 18,373 galaxies.

\subsubsection{Decomposed Samples} \label{Sec: Decomposed AGN Selection}
The ZFOURGE survey is first decomposed into SF and AGN components using the CIGALE software. Each ZFOURGE source has a component AGN and SF luminosity in the CIGALE samples. For a comprehensive explanation of the SED decomposition process, please refer to Section \ref{Sec: CIGALE}. We continue to apply the ZFOURGE survey's \texttt{Use=1} for our analysis, resulting in an initial sample of 37,647 sources for both the AGN and SF samples. Most objects will have a variable combination of AGN and SF that fit into both decomposed samples. We do not remove ZFOURGE sources labelled $L_{IR}<0$ in the SF or AGN samples and instead let CIGALE recalculate the component luminosities. As such, the decomposed samples may have more sources than the ZFOURGE sample. 

For our AGN-specific analysis, we include all sources with a significant AGN fraction ($\mathcal{F}_{AGN}>0.1\ L_{AGN}/L_{IR}$); 21,063 sources. We apply an 80\% bolometric flux cut (calculated from the bolometric SF and AGN luminosities, respectively) to both decomposed samples as was performed with ZFOURGE. After applying each flux cut, the SF and AGN sample is reduced to a final set of 16,850 and 30,059 sources, respectively, spanning $0 \leq z \leq 6$. 

Throughout this work, we use ``decomposed" to refer to the CIGALE SF and AGN luminosities and samples and ``ZFOURGE" to refer to the original ZFOURGE luminosities and dataset. To avoid confusion, the ``bolometric luminosities” referred to in this work are infrared luminosities integrated over rest-frame 8–1000$\mu$m, and do not include emission from X-ray, UV, or radio wavelengths.

\begin{figure}
    \centering
    \includegraphics[width=0.48\textwidth]{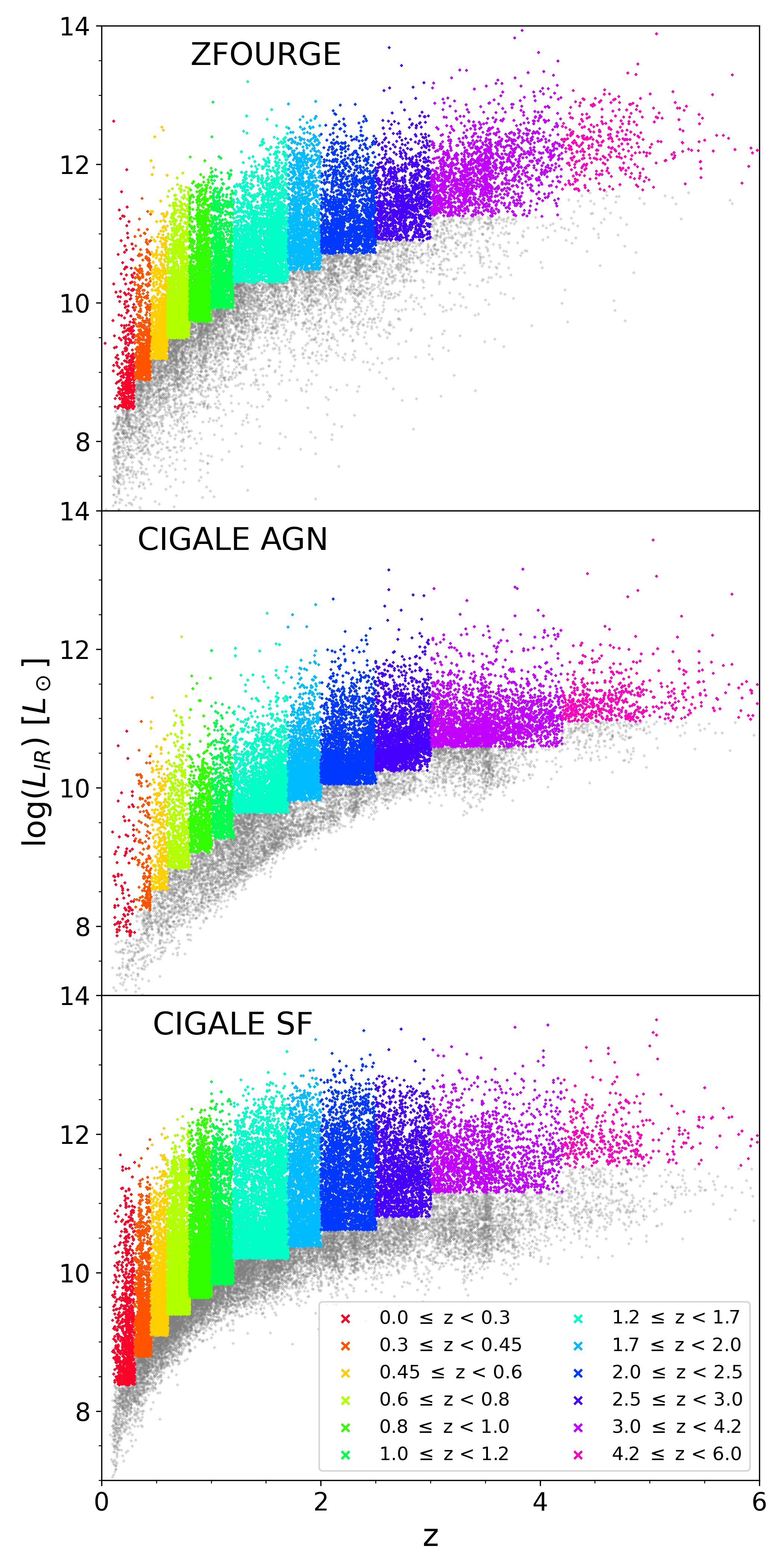}
    \caption{Luminosity-redshift distributions of (top) the ZFOURGE bolometric $8-1000\mu m$ IR luminosity, (middle) the CIGALE AGN luminosity, and (bottom) the CIGALE SF luminosity. Sources are coloured by redshift bin or coloured grey if removed as described in section \ref{Sec: Sample Selection}}
    \label{Fig: ZF Lum vs z}
\end{figure}

Figure \ref{Fig: ZF Lum vs z} shows the reduced luminosity-redshift distribution of ZFOURGE (top), CIGALE AGN (middle), and CIGALE SF (bottom). The jagged boundaries between redshift bins represent the luminosity completeness limit of the respective luminosity function. The luminosity completeness limit is defined as the turnover in the luminosity function.

\section{Decomposing Galaxy SEDs} \label{Sec: CIGALE}
The public version of the ZFOURGE catalogues has utilised \texttt{EAZY} \citep{brammer_eazy_2008} and \texttt{FAST} \citep{kriek_ultra-deep_2009} for parameterising galaxy properties, primarily focusing on photometric redshifts, stellar masses, and SFRs. However, while effective, these methods provide a more generalised view of galaxy properties without entirely disentangling the contributions from different physical components within each galaxy, such as SF regions and AGN. To address this limitation, we employ the SED fitting software CIGALE \citep{boquien_cigale_2019}, which enables the decomposition of observed light from galaxies into distinct components, including SF and AGN activity. This method builds on the work by \cite{cowley_decoupled_2018}, incorporating additional photometric coverage and an updated parameter space to better quantify AGN contributions.

\subsection{CIGALE Methodology and Parameter Space} \label{Sec: CIGALE_Parameters}
CIGALE performs multi-component SED fitting to derive galaxy properties by integrating our photometry from 0.2 $\mu$m to 160 $\mu$m across the CDFS field and up to 24 $\mu$m for COSMOS and UDS. This broader wavelength coverage allows for a more complete and precise decomposition of galaxy light into SF and AGN components. The decomposition uses a range of parameter values (see Table \ref{tab:parameter_space}), allowing for flexible modelling of star formation histories (SFH), dust attenuation, AGN torus contributions, and other factors. We also incorporate the SKIRTOR AGN torus model \citep{stalevski_dust_2016}, which better handles clumpy dust distributions and polar dust extinction, providing an accurate characterisation of AGN emission.

The ZFOURGE sample has the following photometry: IRAC 3.6$\mu m$ (S/N $\geq$ 1): 98.84\%; MIPS 24$\mu m$ (S/N $\geq$ 1): 69.09\%; and Herschel PACS 70-160$\mu m$ (S/N $\geq$ 1): 11.84\%. With each band increasing to 99.47\%, 100\% and 26.19\% at S/N $\geq$ 0 respectively. 

\subsection{Bolometric IR Luminosity Derivation} \label{Sec: IR_Luminosity}
Understanding the total IR energy output of galaxies is essential for tracing both SF and AGN activity, particularly in dusty environments where much of the energy is re-emitted in the IR \citep{fu_decomposing_2010}. By estimating the bolometric IR luminosity, we can gain insights into the contribution of these processes across cosmic time. 

\begin{figure}
    \centering
    \includegraphics[width=0.48\textwidth]{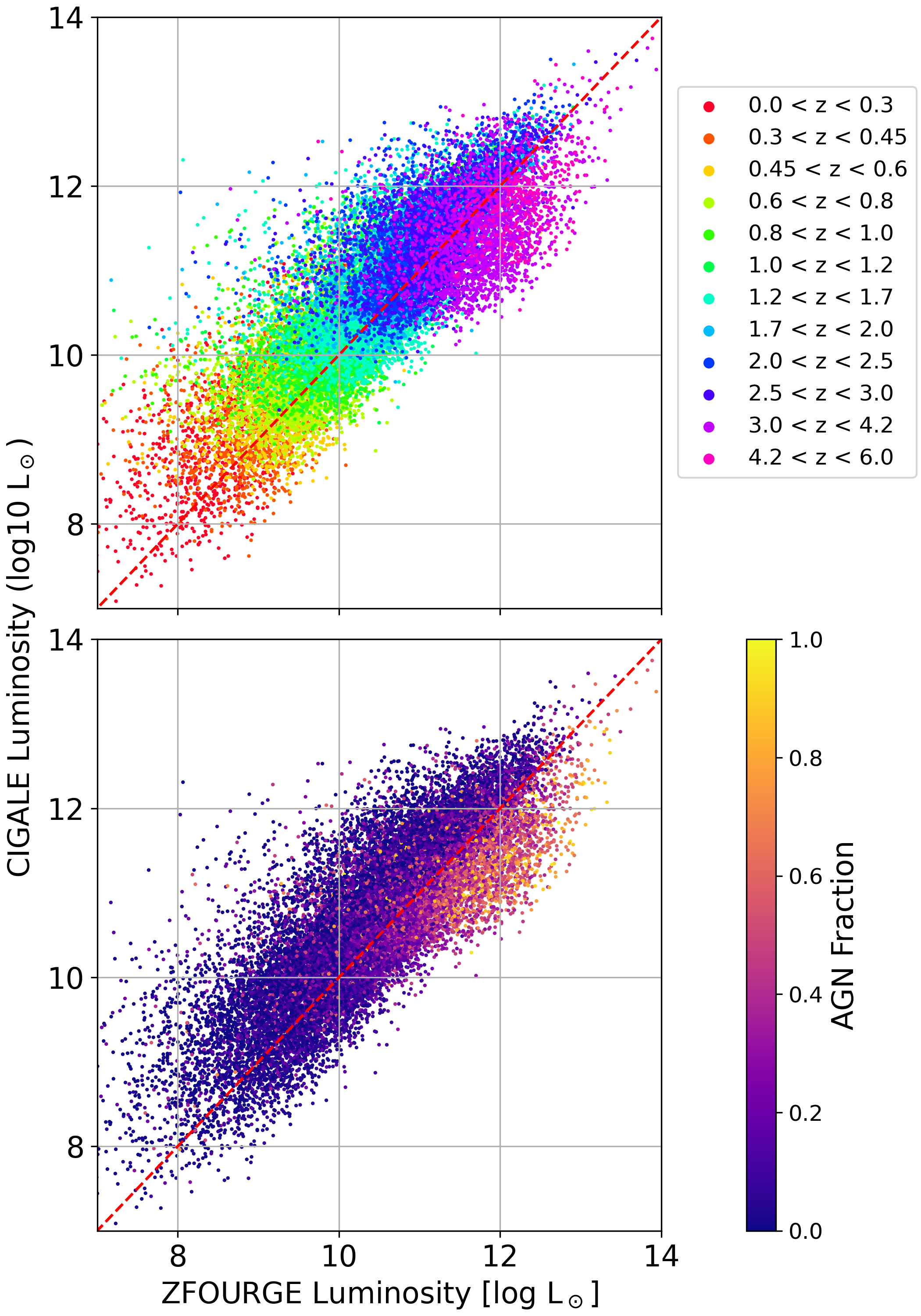}
    \caption{ZFOURGE bolometric 8-1000$\mu$m IR luminosity compared to CIGALE total luminosity. Top: Sources coloured by redshift bin. Bottom: sources coloured by AGN fraction ($\mathcal{F}_{AGN}$). AGN fraction increases with redshift. At $z \geq 3$, the average AGN fraction is greater than 30\%.}
    \label{Fig: LIR vs LIR}
\end{figure}

For the ZFOURGE LF sample, we adopted the approach in \cite{straatman_fourstar_2016} where the averaged \cite{wuyts_fireworks_2008} template was fit to the 24-160 $\mu$m photometry to estimate the total bolometric IR luminosity. However, at $z>3$, we find an offset of 0.27 dex in ZFOURGE luminosity over CIGALE luminosity. This is readily seen in Figure \ref{Fig: LIR vs LIR} (top) as the purple/magenta dots. These sources are also highly AGN-dominated. Because \cite{wuyts_fireworks_2008} only samples the 24-160$\mu m$ range, in conjunction with our dataset hosting only a small fraction of FIR data, our ZFOURGE luminosities at $z>3$ are likely overestimated. To account for this, we add in quadrature to the ZFOURGE $L_{IR}$ an error of 0.27 dex in addition to $z_{phot}$ and $L_{IR}$ errors.

For the CIGALE decomposed SF and AGN LF samples, we derive luminosities through a different approach. CIGALE performs SED decomposition on the entire galaxy emission, using its integrated models to separate the total luminosity into distinct stellar, dust, and AGN components. The stellar and dust luminosities are combined to form the SF component, while the AGN component is derived directly from CIGALE's emission modelling. These distinct approaches are subsequently used to construct the IR luminosity functions of the two samples. 

Figure \ref{Fig: LIR vs LIR} compares the CIGALE decomposed total luminosity to the ZFOURGE total luminosity ($L_{IR}$). In the top panel, galaxies are coloured based on their redshift bin. It can be seen that the brightest galaxies are more likely to exist at higher redshifts. In the bottom panel, galaxies are coloured based on the AGN fraction ($\mathrm{F}_{AGN}$) to the total luminosity derived by CIGALE. Galaxies at higher redshift are brighter and more likely to host a powerful AGN.

\begin{table}[htbp]
    \caption{Parameter space used for SED fitting with CIGALE}
    \label{tab:parameter_space}
    \begin{center}
    \begin{tabular}{ll}
        \toprule
        \textbf{Parameter} & \textbf{Model/Values} \\ 
        \hline
        SFH                 & Delayed SFH $\tau = 1,3,5,7,9,11$ Gyr \\
        Age                 & $0.5, 1, 3, 5, 7, 9, 11$ Gyr \\
        Burst Fraction      & $0.0, 0.01, 0.05, 0.1, 0.15, 0.2, 0.3$ \\
        SSP                 & \cite{bruzual_stellar_2003} \\
        IMF                 & \cite{chabrier_galactic_2003} \\
        Metallicity         & Fixed at 0.02 \\
        Nebular             & \cite{inoue_rest-frame_2011} \\
        Dust Atten.         & \cite{calzetti_dust_2000} $E_{(B-V)} = 0.01, 0.05, 0.1, 0.5$, \\
                            & $1.0, 1.5$ \\
        Dust Emission       & \cite{dale_two-parameter_2014} $\alpha = 1.0, 1.5, 2.0, 2.5, 3.0$ \\
        AGN Model           & SKIRTOR \citep{stalevski_3d_2012, stalevski_dust_2016} \\
        Torus Inclination   & $30^\circ, 70^\circ$ \\
        AGN Fraction        & $0.0, 0.01, 0.1 - 0.9$ (steps of 0.1), 0.99 \\
        Polar Extinction    & SMC $E(B-V) = 0.0, 0.03, 0.1, 0.2, 0.4, 0.6, 1.0, 1.8$ \\
        \botrule
    \end{tabular}
    \end{center}
\end{table}

\subsection{Robustness Tests with Mock Analysis} \label{Sec: Mock_Analysis}
To ensure the reliability of the decomposition process, particularly for faint AGN, we performed a series of robustness tests using CIGALE's built-in mock analysis. These tests evaluate the software's ability to accurately decompose AGN and SF contributions across various redshifts and luminosities, specifically focusing on galaxies with low bolometric IR luminosities. By comparing the input and recovered AGN luminosities from the analysis, we confirmed that our parameter space and methodology are robust, particularly in detecting faint AGN. The mock analysis demonstrated that AGN luminosity was reliably constrained, with Pearson correlation coefficients (PCCs) ranging from 0.969 to 0.973 across all fields. Most sources lay within 0.5 dex of the 1-to-1 line, with mean residuals between $-0.02$ and $0.04$ dex, confirming the robustness of AGN luminosity recovery. These results indicate that our method effectively minimises bias against faint AGN, which are often difficult to detect in traditional analyses.

\subsection{Reliability of $L_{IR}$ Estimates Without FIR Constraints} \label{Sec: FIR Constraints}
ZFOURGE provides deep photometric coverage from the UV through near- and mid-infrared, but coverage at far-infrared (FIR) wavelengths, particularly at 160$\mu m$, is limited (Refer to Section \ref{Sec: CIGALE_Parameters}). Herschel PACS photometry is available for only a subset of sources in the CDFS field. To assess how this limitation affects our ability to recover LIR we performed a targeted mock-based validation using CIGALE.

We selected all galaxies with redshift $z > 2.5$ (a total of 6,940 sources) and compared their estimated $L_{IR}$ to the corresponding mock-recovered values provided by CIGALE. This test does not impose a MIPS SNR cut, ensuring that it captures the uncertainty across the full range of constraints available. Restricting the comparison to galaxies within the range $\log_{10}(L_{IR} [W]) \in [34, 40]$, we find:

\begin{itemize}
    \item Mean $\Delta \log L_{IR}$ = 0.004 dex (no systematic bias),
    \item Standard deviation = 0.27 dex,
    \item Normalized median absolute deviation (NMAD) = 0.14 dex
    \item ~19\% of sources show deviations greater than 0.3 dex.
\end{itemize}

These results indicate that CIGALE can recover IR luminosities with quantified and stable accuracy even for high-z sources lacking strong FIR constraints. This provides a quantitative bound on the uncertainty in $L_{IR}$ used throughout the luminosity function analysis. We add an additional error of 0.27 dex in luminosity to all sources at all redshifts and propagate this uncertainty to the luminosity function and beyond.

For a more comprehensive description of the SED decomposition methodology, see \cite{cowley_decoupled_2018}. Additionally, refer to the CIGALE software paper \citep{boquien_cigale_2019} for detailed information on its decomposition process and parameter optimisation techniques.

\section{Luminosity Functions} \label{Sec: Luminosity Functions}
\subsection{Vmax} \label{Sec: Vmax}

To estimate the LF from our data, we utilise the $1/V_{max}$ method \citep{schmidt_space_1968}. The $1/V_{max}$ method is well-suited for surveys like ZFOURGE, as it does not assume any specific shape for the LF and can easily accommodate galaxies observed across varying depths. It accounts for the maximum observable volume of each galaxy and is given by equation \ref{EQ: 1/Vmax}:

\begin{equation} \label{EQ: 1/Vmax}
    \phi(L,z) = \frac{1}{\Delta \log L}\sum_{i=1}^N \frac{1}{V_{max,i}}
\end{equation}

Where $V_{max}$ represents the maximum co-moving volume of the $i$-th source and $\Delta$ log(L) is the width of the luminosity bin. In practice, to observe the evolution of the LF through cosmic time, the maximum observable volume ($V_{max}$) is calculated for each redshift bin where the upper and lower bounds of the redshift bin limit the volume. Additionally, redshift bins are split into luminosity bins to observe the number density evolution across the different classes of luminosity such as LIRGs (10$^{11} < L_{IR} < 10^{12}\ L_{\odot}$) and ULIRGs ($L_{IR} > 10^{12}\ L_{\odot}$). $V_{max}$ of each galaxy is calculated by taking the maximum comoving volume of the redshift bin the galaxy resides in and subtracting the comoving volume at the beginning of the redshift bin (equation \ref{EQ: Vmax}). We account for the survey area of ZFOURGE (0.1111 degrees$^2$), which normalises the volume probed across the sky (41,253 degrees$^2$). 

\begin{figure*}
    \centering
    \includegraphics[width=\textwidth]{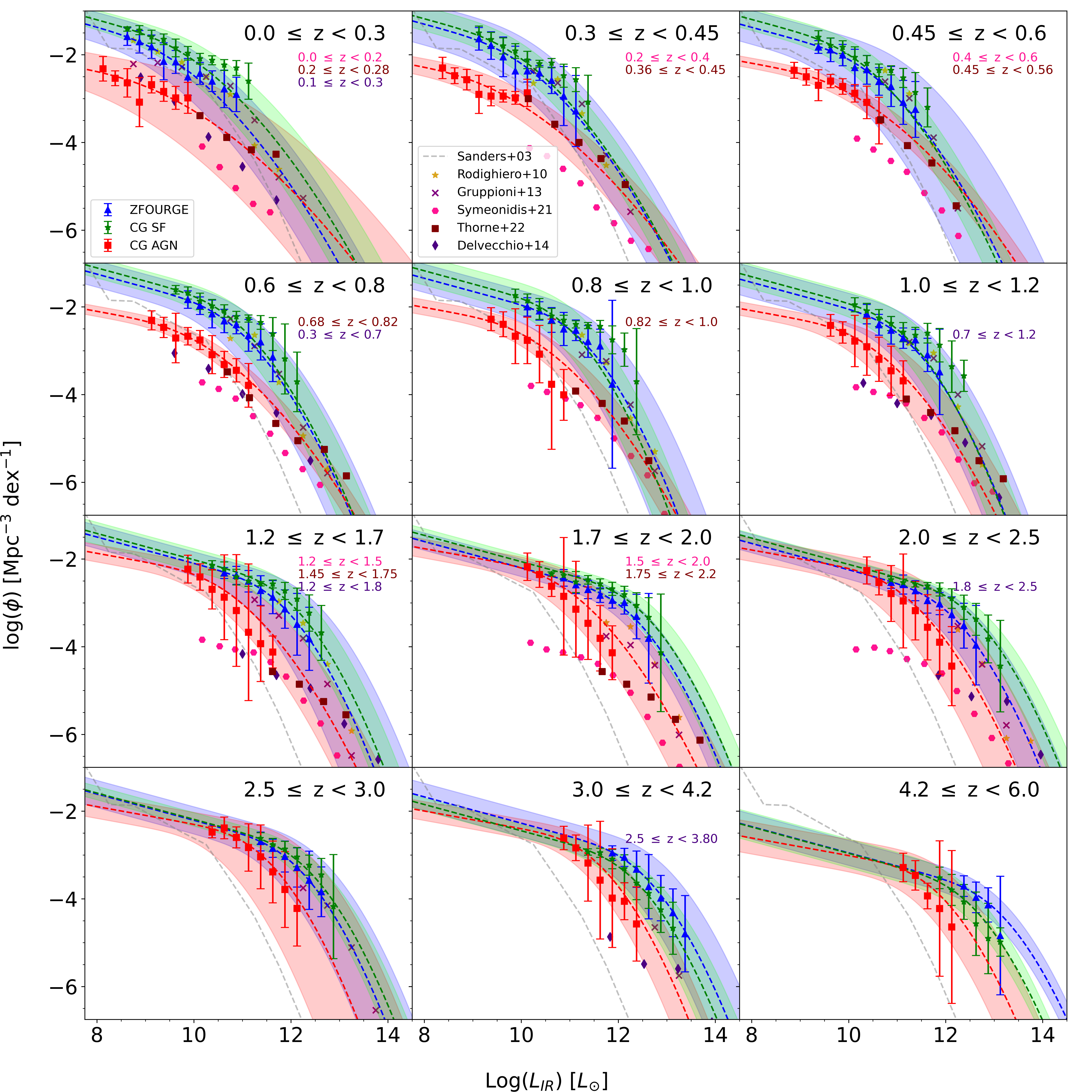}
    \caption{The luminosity functions of major galaxy populations in ZFOURGE and CIGALE calculated using the Vmax method. The dark blue triangles present the ZFOURGE bolometric IR (8-1000$\mu$m) LF. The CIGALE SF and AGN LFs are the green stars and red squares, respectively. The blue, red, and green dashed lines show the best fit Saunders function \citep{saunders_60-mum_1990} to the ZFOURGE, CIGALE AGN, and CIGALE SF, respectively. The shaded regions represent the functional fit errors. The luminosity completeness limit of each redshift bin is where we stop displaying fainter $\phi$ values. Where possible, comparable literature results are also shown. The local \cite{sanders_iras_2003} luminosity function is shown across all redshift bins as the grey dashed line. \cite{rodighiero_mid-_2010} is shown as gold filled stars from $0 < z < 2.5$, \cite{gruppioni_herschel_2013} as purple crosses from $0 < z < 4.2$, \cite{symeonidis_agn_2021} AGN as pink hexagons, \cite{thorne_deep_2022} AGN as maroon squares, \cite{delvecchio_tracing_2014} AGN as indigo diamonds. Differing redshift bins are colour-labelled accordingly. Our ZFOURGE results are consistent with various sources across redshift bins in the literature \citep{caputi_infrared_2007, huang_local_2007, fu_decomposing_2010}}
    \label{Fig: Bolometric IR LF}
\end{figure*}

\begin{equation}
    \label{EQ: Vmax}
    V_{max,i} = \frac{4}{3} \pi \left(D_{max}^3 - D_{min}^3\right) \times \frac{A}{41,253}
\end{equation}

We calculate the maximum ($D_{max}$) and minimum ($D_{min}$) comoving distances for all sources within each redshift bin using the \texttt{FlatLambdaCDM} model from the \texttt{Astropy} Python package \citep{astropy_collaboration_astropy_2022}. The cosmological parameters used are the same as listed in Section \ref{Sec: Intro}. We limit each luminosity bin to a minimum of five sources, or else the luminosity bin is discarded. The relative LF number density $1\sigma$ error values are calculated with:

\begin{equation} \label{EQ: Vmax Error}
    \phi(L,z) = \frac{1}{\Delta \log L}\sqrt{\sum_i \frac{1}{V_{max}^2}}
\end{equation}

We add in quadrature the errors of $z_{phot}$, $L_{IR}$, and FIR deviation from our reduced mock analysis, significantly increasing the error for each luminosity bin.

\begin{table*}
    \begin{center}
    \caption{ZFOURGE bolometric IR (8-1000$\mu$m) LF $\phi$ values.}
    \label{Tab: ZF LF}
    \begin{tabular}{@{}ccccccc@{}}
        \toprule
        $\log_{10}(L_{IR}/L_{\odot})$ & 0.00 $\leq z <$ 0.30 & 0.30 $\leq z <$ 0.45 & 0.45 $\leq z <$ 0.60 & 0.60 $\leq z <$ 0.80 & 0.80 $\leq z <$ 1.00 & 1.00 $\leq z <$ 1.20 \\
        \hline
         8.50 --  8.75 & -1.58 $\pm$ 0.20 & - & - & - & - & - \\
         8.75 --  9.00 & -1.70 $\pm$ 0.26 & - & - & - & - & - \\
         9.00 --  9.25 & -1.82 $\pm$ 0.27 & -1.64 $\pm$ 0.26 & - & - & - & - \\
         9.25 --  9.50 & -2.15 $\pm$ 0.42 & -1.79 $\pm$ 0.31 & -1.82 $\pm$ 0.15 & - & - & - \\
         9.50 --  9.75 & -2.16 $\pm$ 0.26 & -2.05 $\pm$ 0.45 & -1.89 $\pm$ 0.27 & - & - & - \\
         9.75 -- 10.00 & -2.51 $\pm$ 0.42 & -2.37 $\pm$ 0.57 & -2.01 $\pm$ 0.24 & -1.83 $\pm$ 0.20 & - & - \\
        10.00 -- 10.25 & -2.43 $\pm$ 0.25 & -2.37 $\pm$ 0.25 & -2.29 $\pm$ 0.42 & -1.97 $\pm$ 0.20 & -2.00 $\pm$ 0.21 & - \\
        10.25 -- 10.50 & -2.60 $\pm$ 0.24 & -2.41 $\pm$ 0.17 & -2.34 $\pm$ 0.33 & -2.16 $\pm$ 0.32 & -2.09 $\pm$ 0.21 & -2.16 $\pm$ 0.22 \\
        10.50 -- 10.75 & -2.78 $\pm$ 0.43 & -2.58 $\pm$ 0.30 & -2.61 $\pm$ 0.31 & -2.32 $\pm$ 0.27 & -2.30 $\pm$ 0.28 & -2.41 $\pm$ 0.24 \\
        10.75 -- 11.00 & -2.90 $\pm$ 0.39 & -2.94 $\pm$ 0.67 & -2.72 $\pm$ 0.46 & -2.41 $\pm$ 0.21 & -2.50 $\pm$ 0.38 & -2.53 $\pm$ 0.31 \\
        11.00 -- 11.25 & -                & -3.28 $\pm$ 0.81 & -3.09 $\pm$ 0.50 & -2.65 $\pm$ 0.36 & -2.53 $\pm$ 0.24 & -2.71 $\pm$ 0.23 \\
        11.25 -- 11.50 & -                & -                & -3.25 $\pm$ 0.64 & -2.80 $\pm$ 0.41 & -2.79 $\pm$ 0.36 & -2.75 $\pm$ 0.24 \\
        11.50 -- 11.75 & -                & -                & -                & -3.15 $\pm$ 0.56 & -2.89 $\pm$ 0.38 & -3.10 $\pm$ 0.38 \\
        11.75 -- 12.00 & -                & -                & -                & -                & -3.76 $\pm$ 1.91 & -3.48 $\pm$ 0.97 \\
        \hline
        $\log_{10}(L_{IR}/L_{\odot})$ & 1.20 $\leq z <$ 1.70 & 1.70 $\leq z <$ 2.00 & 2.00 $\leq z <$ 2.50 & 2.50 $\leq z <$ 3.00 & 3.00 $\leq z <$ 4.20 & 4.20 $\leq z <$ 6.00  \\
        \hline
        10.50 -- 10.75 & -2.32 $\pm$ 0.13 & - & - & - & - & - \\
        10.75 -- 11.00 & -2.37 $\pm$ 0.21 & -2.43 $\pm$ 0.19 & -2.53 $\pm$ 0.14 & - & - & - \\
        11.00 -- 11.25 & -2.51 $\pm$ 0.24 & -2.59 $\pm$ 0.20 & -2.58 $\pm$ 0.22 & - & - & - \\
        11.25 -- 11.50 & -2.72 $\pm$ 0.34 & -2.70 $\pm$ 0.21 & -2.72 $\pm$ 0.23 & -2.69 $\pm$ 0.21 & - & - \\
        11.50 -- 11.75 & -2.87 $\pm$ 0.38 & -2.83 $\pm$ 0.16 & -2.95 $\pm$ 0.29 & -2.85 $\pm$ 0.21 & - & - \\
        11.75 -- 12.00 & -3.14 $\pm$ 0.44 & -2.94 $\pm$ 0.18 & -3.02 $\pm$ 0.25 & -3.04 $\pm$ 0.35 & -2.96 $\pm$ 0.17 & - \\
        12.00 -- 12.25 & -3.49 $\pm$ 0.71 & -2.99 $\pm$ 0.26 & -3.27 $\pm$ 0.40 & -3.29 $\pm$ 0.45 & -3.05 $\pm$ 0.21 & - \\
        12.25 -- 12.50 & -3.82 $\pm$ 0.72 & -3.31 $\pm$ 0.51 & -3.52 $\pm$ 0.51 & -3.58 $\pm$ 0.66 & -3.32 $\pm$ 0.41 & -3.71 $\pm$ 0.24 \\
        12.50 -- 12.75 & -                & -3.81 $\pm$ 1.02 & -3.97 $\pm$ 0.91 & -3.84 $\pm$ 0.57 & -3.72 $\pm$ 0.73 & -3.97 $\pm$ 0.35 \\
        12.75 -- 13.00 & -                & -                & -                & -                & -3.98 $\pm$ 0.52 & -4.14 $\pm$ 0.39 \\
        13.00 -- 13.25 & -                & -                & -                & -                & -4.32 $\pm$ 0.54 & -4.84 $\pm$ 1.35 \\
        13.25 -- 13.50 & -                & -                & -                & -                & -4.80 $\pm$ 0.87 & -
        \botrule
    \end{tabular}
    \end{center}
    \begin{tabnote}
        {\textbf{Note}: Luminosity bin $\phi$ values are centred.}\tnp
    \end{tabnote}
\end{table*}

\begin{table*}
    \begin{center}
    \caption{CIGALE AGN LF $\phi$ values.}
    \label{Tab: CG AGN LF}
    \begin{tabular}{@{}ccccccc@{}}
        \toprule
        $\log_{10}(L_{IR}/L_{\odot})$ & 0.00 $\leq z <$ 0.30 & 0.30 $\leq z <$ 0.45 & 0.45 $\leq z <$ 0.60 & 0.60 $\leq z <$ 0.80 & 0.80 $\leq z <$ 1.00 & 1.00 $\leq z <$ 1.20 \\
        \hline
         8.00 --  8.25 & -2.32 $\pm$ 0.28 & -                & -                & -                & -                & - \\
         8.25 --  8.50 & -2.54 $\pm$ 0.17 & -2.30 $\pm$ 0.24 & -                & -                & -                & - \\
         8.50 --  8.75 & -2.64 $\pm$ 0.32 & -2.48 $\pm$ 0.18 & -                & -                & -                & - \\
         8.75 --  9.00 & -3.08 $\pm$ 0.56 & -2.57 $\pm$ 0.23 & -2.36 $\pm$ 0.18 & -                & -                & - \\
         9.00 --  9.25 & -2.68 $\pm$ 0.19 & -2.90 $\pm$ 0.43 & -2.50 $\pm$ 0.19 & -2.31 $\pm$ 0.22 & -                & - \\
         9.25 --  9.50 & -2.84 $\pm$ 0.23 & -2.94 $\pm$ 0.22 & -2.70 $\pm$ 0.35 & -2.47 $\pm$ 0.23 & -2.28 $\pm$ 0.26 & - \\
         9.50 --  9.75 & -2.98 $\pm$ 0.26 & -2.94 $\pm$ 0.14 & -2.60 $\pm$ 0.15 & -2.71 $\pm$ 0.56 & -2.40 $\pm$ 0.28 & -2.42 $\pm$ 0.25 \\
         9.75 -- 10.00 & -2.98 $\pm$ 0.35 & -2.98 $\pm$ 0.14 & -2.72 $\pm$ 0.18 & -2.67 $\pm$ 0.20 & -2.67 $\pm$ 0.63 & -2.58 $\pm$ 0.24 \\
        10.00 -- 10.25 & -                & -2.87 $\pm$ 0.31 & -2.88 $\pm$ 0.20 & -2.76 $\pm$ 0.22 & -2.76 $\pm$ 0.53 & -2.78 $\pm$ 0.59 \\
        10.25 -- 10.50 & -                & -                & -3.09 $\pm$ 0.37 & -3.09 $\pm$ 0.43 & -3.08 $\pm$ 0.66 & -2.91 $\pm$ 0.65 \\
        10.50 -- 10.75 & -                & -                & -3.50 $\pm$ 0.53 & -3.31 $\pm$ 0.30 & -3.76 $\pm$ 1.49 & -3.20 $\pm$ 0.50 \\
        10.75 -- 11.00 & -                & -                & -                & -3.45 $\pm$ 0.27 & -4.01 $\pm$ 0.58 & -3.46 $\pm$ 0.56 \\
        11.00 -- 11.25 & -                & -                & -                & -3.79 $\pm$ 0.50 & -                & -3.69 $\pm$ 0.46 \\
        \hline
        $\log_{10}(L_{IR}/L_{\odot})$ & 1.20 $\leq z <$ 1.70 & 1.70 $\leq z <$ 2.00 & 2.00 $\leq z <$ 2.50 & 2.50 $\leq z <$ 3.00 & 3.00 $\leq z <$ 4.20 & 4.20 $\leq z <$ 6.00  \\
        \hline
         9.75 -- 10.00 & -2.23 $\pm$ 0.32 & -                & -                & -                & -                & - \\
        10.00 -- 10.25 & -2.41 $\pm$ 0.30 & -2.18 $\pm$ 0.31 & -                & -                & -                & - \\
        10.25 -- 10.50 & -2.69 $\pm$ 0.45 & -2.35 $\pm$ 0.29 & -2.26 $\pm$ 0.31 & -2.48 $\pm$ 0.13 & -                & - \\
        10.50 -- 10.75 & -2.87 $\pm$ 0.97 & -2.62 $\pm$ 0.23 & -2.54 $\pm$ 0.28 & -2.38 $\pm$ 0.25 & -                & - \\
        10.75 -- 11.00 & -3.18 $\pm$ 1.27 & -2.85 $\pm$ 1.34 & -2.79 $\pm$ 0.64 & -2.60 $\pm$ 0.24 & -2.61 $\pm$ 0.27 & - \\
        11.00 -- 11.25 & -3.67 $\pm$ 1.56 & -3.15 $\pm$ 1.09 & -2.96 $\pm$ 1.07 & -3.83 $\pm$ 0.54 & -2.84 $\pm$ 0.29 & - \\
        11.25 -- 11.50 & -3.93 $\pm$ 0.86 & -3.47 $\pm$ 0.83 & -3.18 $\pm$ 0.67 & -3.04 $\pm$ 0.73 & -3.19 $\pm$ 0.87 & -3.29 $\pm$ 0.33 \\
        11.50 -- 11.75 & -4.12 $\pm$ 0.36 & -3.81 $\pm$ 0.75 & -3.56 $\pm$ 0.74 & -3.39 $\pm$ 0.70 & -3.57 $\pm$ 1.34 & -3.47 $\pm$ 0.34 \\
        11.75 -- 12.00 & -                & -4.14 $\pm$ 0.61 & -3.90 $\pm$ 0.73 & -3.79 $\pm$ 0.86 & -3.98 $\pm$ 1.13 & -3.94 $\pm$ 0.29 \\
        12.00 -- 12.25 & -                & -                & -4.44 $\pm$ 0.90 & -4.22 $\pm$ 0.86 & -4.06 $\pm$ 0.42 & -4.22 $\pm$ 1.54 \\
        12.25 -- 12.50 & -                & -                & -                & -                & -4.57 $\pm$ 0.85 & -4.64 $\pm$ 1.74
        \botrule
    \end{tabular}
    \end{center}
    \begin{tabnote}
        {\textbf{Note}: Luminosity bin $\phi$ values are centred.}\tnp
    \end{tabnote}
\end{table*}

\begin{table*}
    \begin{center}
    \caption{CIGALE SF LF $\phi$ values.}
    \label{Tab: CG SF LF}
    \begin{tabular}{@{}ccccccc@{}}
        \toprule
        $\log_{10}(L_{IR}/L_{\odot})$ & 0.00 $\leq z <$ 0.30 & 0.30 $\leq z <$ 0.45 & 0.45 $\leq z <$ 0.60 & 0.60 $\leq z <$ 0.80 & 0.80 $\leq z <$ 1.00 & 1.00 $\leq z <$ 1.20 \\
        \hline
         8.50 --  8.75 & -1.41 $\pm$ 0.05 & - & - & - & - & - \\
         8.75 --  9.00 & -1.44 $\pm$ 0.10 & - & - & - & - & - \\
         9.00 --  9.25 & -1.59 $\pm$ 0.15 & -1.50 $\pm$ 0.16 & - & - & - & - \\
         9.25 --  9.50 & -1.65 $\pm$ 0.12 & -1.65 $\pm$ 0.16 & -1.60 $\pm$ 0.15 & - & - & - \\
         9.50 --  9.75 & -1.84 $\pm$ 0.20 & -1.80 $\pm$ 0.18 & -1.79 $\pm$ 0.17 & -1.61 $\pm$ 0.10 & - & - \\
         9.75 -- 10.00 & -1.97 $\pm$ 0.15 & -1.91 $\pm$ 0.17 & -1.83 $\pm$ 0.14 & -1.68 $\pm$ 0.14 & -1.74 $\pm$ 0.15 & - \\
        10.00 -- 10.25 & -2.09 $\pm$ 0.12 & -2.12 $\pm$ 0.20 & -2.06 $\pm$ 0.24 & -1.87 $\pm$ 0.18 & -1.86 $\pm$ 0.16 & -1.95 $\pm$ 0.16 \\
        10.25 -- 10.50 & -2.13 $\pm$ 0.10 & -2.20 $\pm$ 0.08 & -2.20 $\pm$ 0.18 & -1.98 $\pm$ 0.15 & -2.05 $\pm$ 0.24 & -2.09 $\pm$ 0.17 \\
        10.50 -- 10.75 & -2.23 $\pm$ 0.12 & -2.19 $\pm$ 0.11 & -2.36 $\pm$ 0.14 & -2.09 $\pm$ 0.14 & -2.21 $\pm$ 0.17 & -2.25 $\pm$ 0.19 \\
        10.75 -- 11.00 & -2.30 $\pm$ 0.17 & -2.48 $\pm$ 0.26 & -2.36 $\pm$ 0.10 & -2.22 $\pm$ 0.12 & -2.30 $\pm$ 0.11 & -2.38 $\pm$ 0.16 \\
        11.00 -- 11.25 & -2.60 $\pm$ 0.41 & -2.57 $\pm$ 0.27 & -2.53 $\pm$ 0.22 & -2.27 $\pm$ 0.08 & -2.40 $\pm$ 0.08 & -2.58 $\pm$ 0.22 \\
        11.25 -- 11.50 & -                & -3.08 $\pm$ 0.61 & -2.86 $\pm$ 0.45 & -2.36 $\pm$ 0.16 & -2.43 $\pm$ 0.07 & -2.65 $\pm$ 0.18 \\
        11.50 -- 11.75 & -                & -                & -3.20 $\pm$ 0.44 & -2.62 $\pm$ 0.38 & -2.44 $\pm$ 0.13 & -2.60 $\pm$ 0.22 \\
        11.75 -- 12.00 & -                & -                & -                & -3.19 $\pm$ 0.80 & -2.75 $\pm$ 0.32 & -2.87 $\pm$ 0.38 \\
        12.00 -- 12.25 & -                & -                & -                & -3.71 $\pm$ 0.68 & -2.98 $\pm$ 0.38 & -3.36 $\pm$ 0.59 \\
        12.25 -- 12.50 & -                & -                & -                & -                & -3.71 $\pm$ 1.21 & -3.57 $\pm$ 0.36 \\
        \hline
        $\log_{10}(L_{IR}/L_{\odot})$ & 1.20 $\leq z <$ 1.70 & 1.70 $\leq z <$ 2.00 & 2.00 $\leq z <$ 2.50 & 2.50 $\leq z <$ 3.00 & 3.00 $\leq z <$ 4.20 & 4.20 $\leq z <$ 6.00  \\
        \hline
        10.25 -- 10.50 & -2.15 $\pm$ 0.11 & - & - & - & - & - \\
        10.50 -- 10.75 & -2.23 $\pm$ 0.16 & -2.35 $\pm$ 0.05 & - & - & - & - \\
        10.75 -- 11.00 & -2.41 $\pm$ 0.18 & -2.38 $\pm$ 0.09 & -2.47 $\pm$ 0.07 & - & - & - \\
        11.00 -- 11.25 & -2.51 $\pm$ 0.11 & -2.53 $\pm$ 0.24 & -2.54 $\pm$ 0.08 & - & - & - \\
        11.25 -- 11.50 & -2.56 $\pm$ 0.05 & -2.47 $\pm$ 0.10 & -2.57 $\pm$ 0.06 & -2.62 $\pm$ 0.13 & -2.93 $\pm$ 0.07 & - \\
        11.50 -- 11.75 & -2.57 $\pm$ 0.09 & -2.57 $\pm$ 0.12 & -2.61 $\pm$ 0.08 & -2.77 $\pm$ 0.15 & -2.95 $\pm$ 0.10 & - \\
        11.75 -- 12.00 & -2.73 $\pm$ 0.19 & -2.68 $\pm$ 0.15 & -2.69 $\pm$ 0.15 & -2.87 $\pm$ 0.18 & -3.09 $\pm$ 0.22 & -3.52 $\pm$ 0.25 \\
        12.00 -- 12.25 & -2.92 $\pm$ 0.25 & -2.81 $\pm$ 0.19 & -2.90 $\pm$ 0.22 & -3.05 $\pm$ 0.22 & -3.31 $\pm$ 0.32 & -3.80 $\pm$ 0.35 \\
        12.25 -- 12.50 & -3.23 $\pm$ 0.41 & -3.06 $\pm$ 0.29 & -3.10 $\pm$ 0.28 & -3.23 $\pm$ 0.23 & -3.64 $\pm$ 0.38 & -4.07 $\pm$ 0.40 \\
        12.50 -- 12.75 & -3.68 $\pm$ 0.62 & -3.33 $\pm$ 0.45 & -3.38 $\pm$ 0.36 & -3.46 $\pm$ 0.38 & -3.88 $\pm$ 0.34 & -4.57 $\pm$ 0.72 \\
        12.75 -- 13.00 & -                & -4.14 $\pm$ 1.34 & -3.82 $\pm$ 0.55 & -4.18 $\pm$ 1.19 & -4.25 $\pm$ 0.52 & -4.91 $\pm$ 0.80 \\
        13.00 -- 13.25 & -                & -                & -4.44 $\pm$ 1.04 & -                & -4.67 $\pm$ 0.59 & -4.98 $\pm$ 0.32 
        \botrule
    \end{tabular}
    \end{center}
    \begin{tabnote}
        {\textbf{Note}: Luminosity bin $\phi$ values are centred.}\tnp
    \end{tabnote}
\end{table*}

\subsection{Fitting Functions}
We first construct the bolometric IR LF using the ZFOURGE dataset. Then, using CIGALE, we decompose the luminosity into contributions from SF regions and AGN, allowing us to investigate the evolution of these components separately.

To model LFs, one of the most widely used methods is the Schechter function \citep{schechter_analytic_1976}. However, the bright end slope of the Schechter function cannot be independently varied to better fit a dataset. We make use of a modified Schechter function known as the Saunders function (\citealp{saunders_60-mum_1990}; equation. \ref{EQ: Saunders Function}) to fit our LFs:

\begin{equation} 
    \varphi(L) = \varphi^* \left(\frac{L}{L^*}\right)^{1-\alpha} \exp\left[-\frac{1}{2\sigma^2}\log_{10}^2\left(1+\frac{L}{L^*}\right)\right]
    \label{EQ: Saunders Function}
\end{equation}

Where $\varphi(L)$ is the number of galaxies per unit volume (number density), $\varphi^*$ is the characteristic normalisation factor, $L$ is the bolometric IR (8-1000$\mu$m) luminosity, $L^*$ is the characteristic luminosity, $\alpha$ is the faint end slope, and $\sigma$ is the bright end slope

Our deep ZFOURGE data probes to fainter luminosities than often seen in the literature (e.g. \citealp{rodighiero_mid-_2010, gruppioni_herschel_2013}), thus better constraining the faint end of the LF. However, as ZFOURGE is designed to probe deeper into the universe, we lack brighter galaxies at lower redshifts. 

\section{Discussion} \label{Sec: Discussion}
\subsection{Bolometric IR LF} \label{Sec: Bolometric IR LF}

Figure \ref{Fig: Bolometric IR LF} presents the bolometric IR LFs derived from the ZFOURGE and CIGALE samples. This comparison allows us to explore the evolution of total IR emission and the individual contributions from SF and AGN across twelve redshift bins from $0 \leq z < 6$. By comparing the LF from ZFOURGE with the decomposed SF and AGN LFs from CIGALE, we aim to better understand the distinct roles of SF and AGN activity in galaxy evolution over cosmic time. To model these LFs, we employ \texttt{scipy.optimize.curve\_fit} \citep{virtanen_scipy_2020} which performs non-linear least squares fitting, optimising model parameters by minimising the difference between the model and observed data. 

Relative 1$\sigma$ parameter dispersion errors were calculated using \texttt{np.sqrt(np.diag(pcov))} from \texttt{NumPy} \citep{harris_array_2020}, which relates the covariance of the best-fit parameters. Additionally, the fitting routine is applied to a ``worst-case" upper and lower luminosity based on the accrued errors of redshift, $L_{IR}$, and FIR availability. The errors from these ``worst cases" are then added in quadrature to the uncertainty of the parameters. The shaded regions shown in Figure \ref{Fig: Bolometric IR LF} represent the combined uncertainties, including observational errors (photometric redshifts, LIR, and FIR limitations) and fitting uncertainties (including covariance between fitted LF parameters). These regions thus reflect the plausible range in the LF shape given the data.

It is important to note that the LFs derived in this work apply to the population of galaxies detectable within a near-infrared selection framework, as discussed in Section \ref{Sec: Galaxy LF Selection}. As such, they may underrepresent the most heavily obscured galaxies and AGN, and likely constitute a lower limit on the true space density of dust-obscured systems. This limitation is expected to primarily affect the faint end of the luminosity function, where obscured but intrinsically faint galaxies are most likely to be missed. However, the bright end, which is dominated by luminous systems, is less sensitive to this bias and should remain largely representative. As such, our luminosity functions should be regarded as conservative lower limits, particularly at lower luminosities. Again, as mentioned in Sections \ref{Sec: IR_Luminosity} and \ref{Sec: FIR Constraints}, we include and account for FIR errors to capture the true shape of each LF inside our uncertainty.

\subsubsection{ZFOURGE} \label{Sec: ZF Total Discussion}
Focusing first on the ZFOURGE data, we compare our results with \cite{rodighiero_mid-_2010} and \cite{gruppioni_herschel_2013}. We also compare our results with \cite{huang_local_2007, caputi_infrared_2007, fu_decomposing_2010} but avoid cluttering figure \ref{Fig: Bolometric IR LF} with these disjointed redshift bins from the literature. Across all redshift bins (except the most local), we consistently see that the ZFOURGE number density ($\phi$) values in blue extend much fainter than the rest of the literature, showcasing ZFOURGE's ability to probe to fainter luminosities. However, there remains room to improve the constraints at the faint end of the ZFOURGE LF. Extending the analysis by an additional order of magnitude fainter in each redshift bin would significantly enhance our ability to constrain the faint-end slope.

We do not compare the CIGALE Total LF to the ZFOURGE LF. This is because the CIGALE Total LF is practically identical to the CIGALE SF LF and would clutter Figure \ref{Fig: Bolometric IR LF}. Instead, we present the LFs of ZFOURGE, CIGALE Total, and CIGALE components in Appendix A at the end of this article.

\begin{figure}
    \centering
    \includegraphics[width=0.48\textwidth]{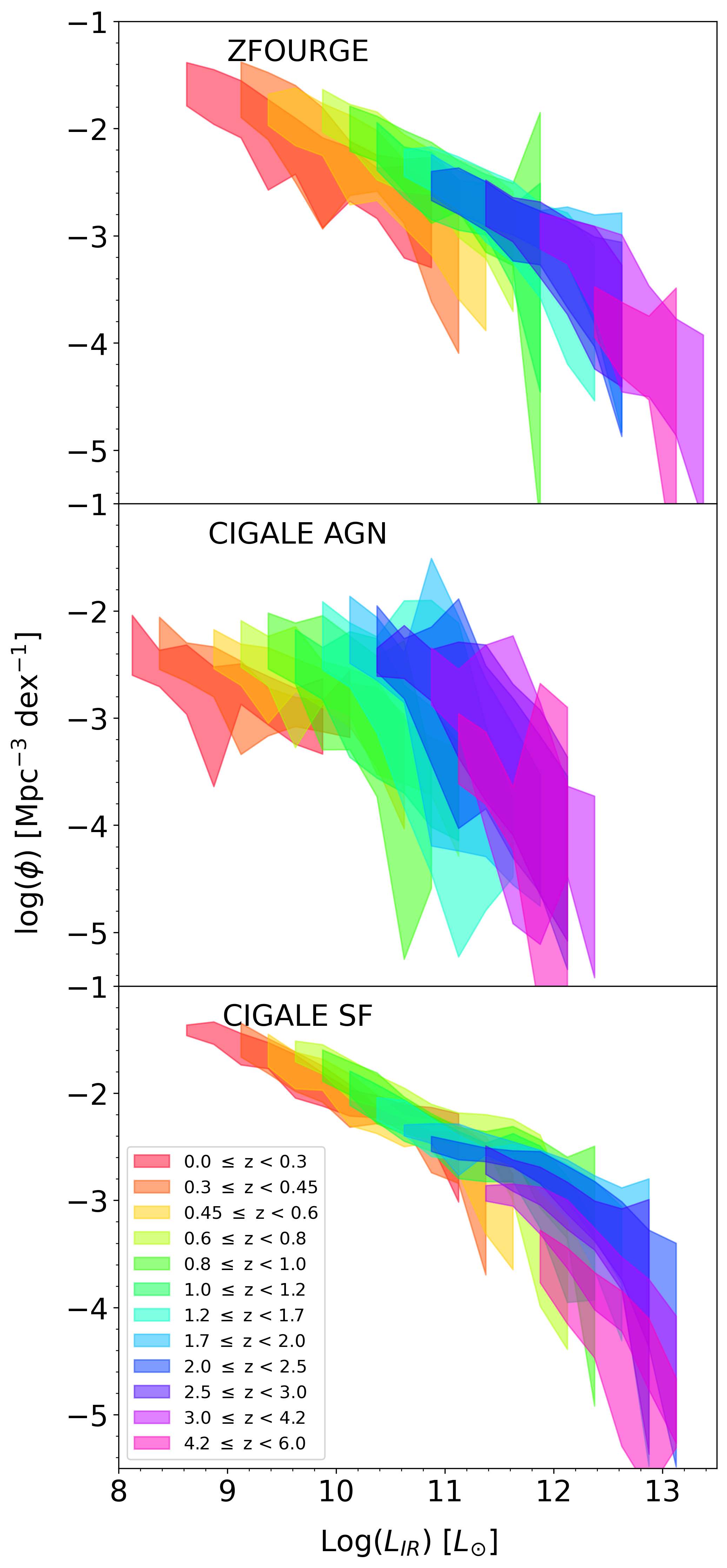}
    \caption{Combined evolution of the ZFOURGE (top), CIGALE AGN (middle), and CIGALE SF (bottom) luminosity functions shown in figure \ref{Fig: Bolometric IR LF}. The redshift evolution of each binned LF is easier to visualise. Data points are shaded between the uncertainties and coloured by redshift bin.}
    \label{Fig: LF Filled}
\end{figure}

Our ZFOURGE results agree very well with the literature in all bins except $1.7 \leq z < 2$ and $3.0 \leq z < 4.2$. In these redshift bins, ZFOURGE results show $\phi$ values $\approx$ 0.2 --- 0.5 dex higher across all luminosity bins when compared to \cite{rodighiero_mid-_2010} and \cite{gruppioni_herschel_2013}. We posit that ZFOURGE is detecting fainter sources in these redshift bins than previously observed. From $1.7 \leq z < 2$, both \cite{rodighiero_mid-_2010} and \cite{gruppioni_herschel_2013} show a drop in their faintest luminosity bins. The fact that both \cite{gruppioni_herschel_2013} and \cite{rodighiero_mid-_2010} see a drop from $1.7 \leq z < 2$ is intriguing. This issue does not appear in neighbouring redshift bins or other redshift bins. \cite{rodighiero_mid-_2010} utilises multiwavelength Spitzer observations, whereas \cite{gruppioni_herschel_2013} uses Herschel/PACS data to estimate the total IR LF. Given that this exists across multiple surveys and instruments, it remains to be seen why a drop in the $1.7 \leq z < 2$ redshift bin exists. Our ZFOURGE results, on the other hand, do not show this drop. Luminosity bins in our final redshift bin $4.2 \leq z < 6.0$ (figure \ref{Fig: Bolometric IR LF}) show a drop along the faint end slop of our LF. Therefore, our final redshift bin $4.2 \leq z < 6.0$ is likely incomplete and should be taken as a lower limit.

\subsubsection{CIGALE AGN}
For the CIGALE AGN, we compare our results with \cite{delvecchio_tracing_2014, symeonidis_agn_2021} and \cite{thorne_deep_2022}. \cite{symeonidis_agn_2021} derives their IR AGN $\phi$ values from the hard X-ray LFs by \cite{aird_evolution_2015}. \cite{thorne_deep_2022}, who performs similar work to this analysis, uses the SED fitting code \texttt{ProSpect} \citep{leja_deriving_2017, robotham_prospect_2020} to decompose the bolometric IR LF and recover the pure AGN component to the LF. We fit the Saunders function in red and uncertainties to our CIGALE decomposed AGN LF. We include \cite{thorne_deep_2022} AGN in our fitting process as we do not have comparatively bright AGN to constrain the bright end of the LF. 

In the first few redshift bins from $0 \leq z < 0.8$, our CIGALE AGN LF and the Saunders function fits are generally consistent with the results in the literature. However, the literature LFs tend to flatten considerably at higher redshifts and fainter luminosities. In contrast, our CIGALE AGN LFs do not flatten and instead continue to rise, suggesting that CIGALE SED decomposition is effective in isolating and recovering the AGN contribution to luminosities as faint as $10^8$ $L_{\odot}$. Furthermore, we probe fainter than \cite{thorne_deep_2022} whose faintest luminosity bins are never less than $10^{10}$ $L_{\odot}$ (when accounting for their completeness limits). Although our CIGALE AGN LF lacks $\phi$ values at the bright end, necessitating the use of \cite{thorne_deep_2022} $\phi$ values to constrain our Saunders function fits, CIGALE's ability to extend the LF to such faint luminosities provides crucial insights into the AGN population at higher redshifts. 

The ZFOURGE LF, especially at the bright end, is elevated above the CIGALE AGN LF across all redshift bins. This may result from AGN contributions that are not captured in the decomposed AGN LF due to limitations in FIR coverage and SED constraints at high luminosities. Additionally, we have shown the \cite{wuyts_fireworks_2008} method (in Section \ref{Sec: IR Density}) likely overestimates the IR luminosity at $z>3$. This is visually represented by Figure \ref{Fig: LIR vs LIR} (bottom). Furthermore, there are many fewer and fainter CIGALE AGN sources to work with at this redshift range. This reduction reflects not only the intrinsic scarcity of luminous AGN at high redshift, but also the stricter quality criteria applied during CIGALE SED fitting. As a result, the CIGALE AGN LF at the bright end may underrepresent some systems that are included in the ZFOURGE total LF, further contributing to the observed discrepancy.

Because the IR AGN identified by \cite{symeonidis_agn_2021} are derived from the hard X-ray LF presented by \cite{aird_evolution_2015}, the observed flattening at fainter luminosities is almost certainly due to X-ray emission not identifying the obscured faint AGN population. In contrast, SED fitting, as applied in our study, allows us to recover these faint AGN, providing a more complete picture of the AGN population \citep{gruppioni_modelling_2011, brown_infrared_2019, thorne_deep_2022}. Although not shown in Figure \ref{Fig: Bolometric IR LF}, the combined type-1 and type-2 AGN from \cite{symeonidis_agn_2021} show elevated number densities, particularly in the $1.2 \leq z < 2.5$ range, and align well with our CIGALE AGN results. This strong agreement underscores the robustness of our approach in isolating obscured faint AGN, especially at higher redshifts.

As the CIGALE AGN LF evolves with redshift, the faint end approaches the number density values of the CIGALE SF LF (discussed in the following subsection) at $0 \leq z < 2.5$ and nearly surpasses them at $z > 2.5$. This trend aligns with the well-known peak of the cosmic SF of galaxies above $z=2$ \citep{madau_cosmic_2014}. Conversely, the AGN fraction increases with redshift and $L_{IR}$ as noted by \cite{symeonidis_agn_2021, thorne_deep_2022} and references therein. Although our results do not yet show AGN number densities overtaking those of SF galaxies, future studies probing higher redshifts will likely reveal this transition, reflecting the dominance of AGN activity in the extremely early universe.

\subsubsection{CIGALE SF}
As seen in Figure \ref{Fig: Bolometric IR LF}, the CIGALE SF LF is elevated above the ZFOURGE and comparable literature LFs at fainter luminosities. We see a tight relationship between CIGALE SF and ZFOURGE LFs, with CIGALE SF slightly elevated above ZFOURGE in all but the highest redshift bins where AGN activity increases. Several factors could contribute to this result. Work by \cite{wu_mid-infrared_2011} has shown that the UV and optical wavelengths follow a Schechter function closely. In contrast, the IR wavelengths have a shallower exponential, which is inconsistent with a Schechter function \citep{symeonidis_what_2019}. Although our results in Figure \ref{Fig: Bolometric IR LF} show fewer luminosity bins at the bright end, we argue that CIGALE accurately isolates the SF fraction and AGN contribution to galaxy emission. 


As expected, the bright-end slope of a Schechter function is too steep to accurately describe the ZFOURGE LF \citep{wu_mid-infrared_2011}, in agreement with the literature \citep{rodighiero_mid-_2010, gruppioni_herschel_2013, symeonidis_what_2019}. Even after removing AGN-identified galaxies (552, \citealp{cowley_zfourge_2016}) and rerunning the analysis, the ZFOURGE LFs do not show an improved Schechter function fit as predicted by \cite{fu_decomposing_2010, wu_mid-infrared_2011}. The most likely reason is that \cite{cowley_zfourge_2016} only identifies the most AGN-dominated sources, leaving fainter-luminosity AGN undetected. AGN activity and SFR are tightly coupled (\citealp{alexander_what_2012} and references within), with both AGN activity and SF likely happening at the same time or offset from each other \citep{cowley_decoupled_2018}. At higher redshifts ($z > 2$), it becomes increasingly essential to disentangle AGN and SF components of galaxy emission to model galaxy evolution accurately.

\subsubsection{Combined Evolution}
In Figure \ref{Fig: LF Filled}, we show the combined evolution of the IR (8-1000$\mu$m) LFs introduced in Figure \ref{Fig: Bolometric IR LF}. The LF is filled between the uncertainty bounds. With this figure, it is easier to see the evolution of the LF across luminosity and redshift. A clear declining density trend is seen with increasing luminosity and redshift.

This result is significant because it highlights how the relative contributions of SF and AGN activity evolve. Both SF and AGN number densities increase with luminosity as the universe ages towards the present day. This suggests a tentative downsizing effect, which we explore further in Section \ref{Sec: Class Density}. The decline in the LF with increasing luminosity and redshift suggests that the early universe contained fewer luminous galaxies, implying lower overall SF and AGN activity. As we move towards the present day, the rising number density of bright galaxies in the LF (until the lowest redshift bins) reflects the growth and evolution of galaxies and their central SMBHs, with an increase in both SF and AGN contributions.

\subsection{Parameter Evolution} \label{Sec: Parameter Evolution}
\begin{figure}[h]
    \centering
    \includegraphics[width=0.48\textwidth]{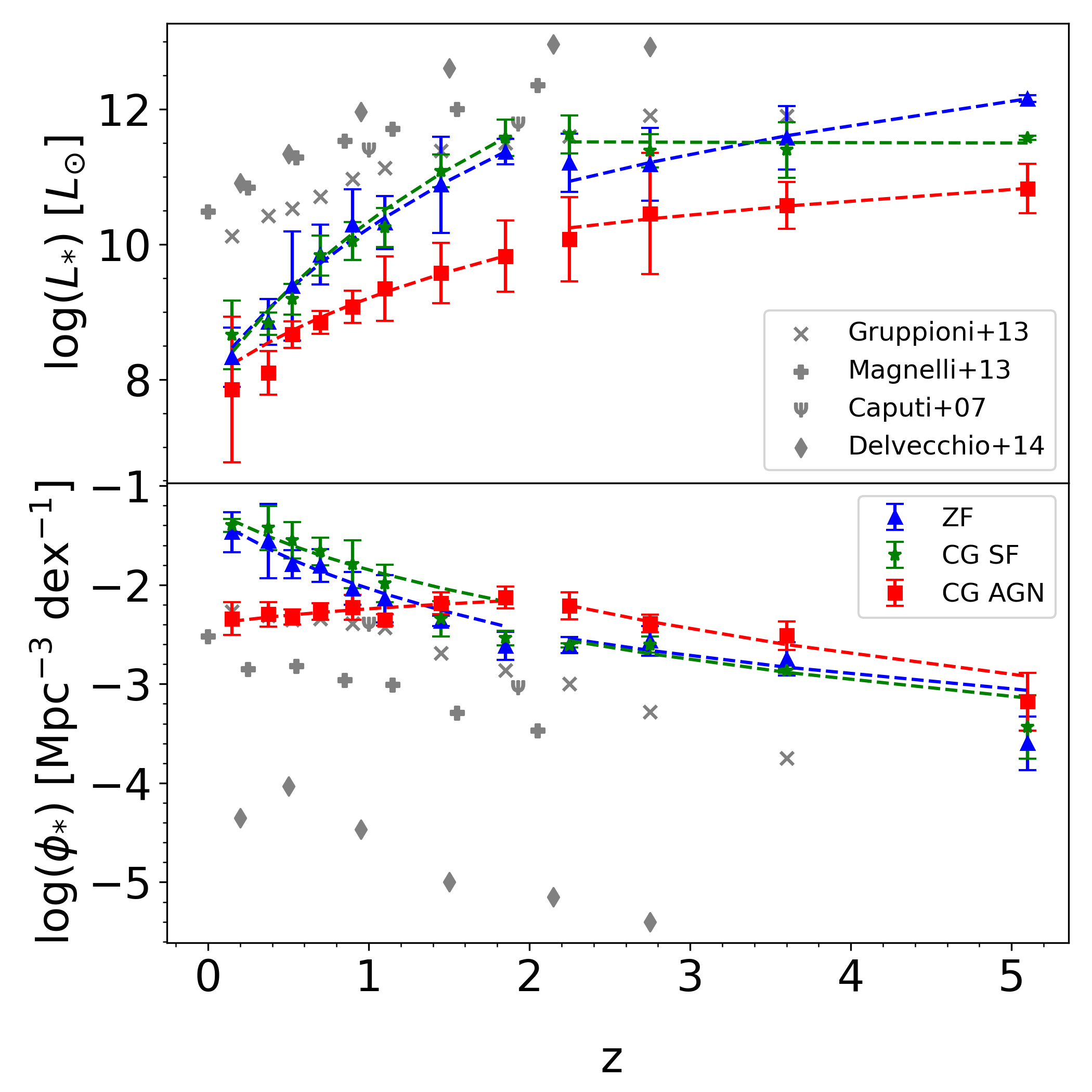}
    \caption{Best fitting parameters and uncertainties to our luminosity functions. Top: $L^{*}$ evolution. Bottom: $\phi^{*}$ evolution. Blue triangles represent the ZFOURGE. Red squares represent the CIGALE AGN and green stars the CIGALE SF population. Dashed lines represent the $\propto(1+z)^k$ evolution. We compare our results to the relevant literature, which is coloured grey. \cite{gruppioni_herschel_2013} crosses, \cite{magnelli_deepest_2013} pluses, and \cite{caputi_infrared_2007} uppercase $\Psi$'s.}
    \label{Fig: Param Evo}
\end{figure}

In this section, we discuss the parameter evolution of the Saunders luminosity function fits in Figure \ref{Fig: Bolometric IR LF}. The evolution of $\phi^{*}$ and $L^{*}$ across redshift is presented in Figure \ref{Fig: Param Evo} with values and errors in Table \ref{Tab: Param Evo}. The best fitting parameter values were calculated following the same fitting routine outlined in Section \ref{Sec: Bolometric IR LF}. As discussed in section \ref{Sec: Bolometric IR LF}, our final redshift bin likely suffers from incompleteness. However, ZFOURGE's ability to probe fainter luminosities becomes advantageous at higher redshifts, providing more reliable constraints on the LF parameters.

To better constrain the faint end slope of all LFs presented in this work, we fix $\alpha=1.3$ across all redshift bins. This result differs from the literature where \cite{rodighiero_mid-_2010, gruppioni_herschel_2013} fix $\alpha=1.2$ whereas \cite{fu_decomposing_2010} leaves $\alpha$ as a free fitting parameter found to be $\alpha=1.46$ (respective to our Saunders fitting function). This is because our LFs extend much further at the faint end. However, as we lack luminosity bins along the bright end of the LF, we fix the bright end $\sigma$ to values that fit best to the available literature. The faint end slopes of CIGALE LFs agree with the literature. Due to the absence of bright-end AGN luminosity bins, we incorporate data from \cite{thorne_deep_2022} to help constrain the AGN LF fitting process.

\begin{table}[h]
    \caption{Best-fit and fixed Saunders parameters for each LF across different redshift bins.}
    \label{Tab: Param Evo}
    \begin{center}
    \begin{tabular}{ccccc}
        \toprule
        $z$ & log$(L^{*})$ & log$(\phi^{*})$ & $\alpha$ & $\sigma$ \\
        \hline
        ZFOURGE \\
        \hline
        0.00 $\leq z <$ 0.30 &  8.33 $\pm$ 0.44 & -1.47 $\pm$ 0.20 & 1.3 & 1.1 \\
        0.30 $\leq z <$ 0.45 &  8.85 $\pm$ 0.34 & -1.56 $\pm$ 0.37 & 1.3 & 1.0 \\
        0.45 $\leq z <$ 0.60 &  9.38 $\pm$ 0.81 & -1.79 $\pm$ 0.14 & 1.3 & 0.9 \\
        0.60 $\leq z <$ 0.80 &  9.84 $\pm$ 0.44 & -1.81 $\pm$ 0.17 & 1.3 & 0.8 \\
        0.80 $\leq z <$ 1.00 & 10.29 $\pm$ 0.52 & -2.04 $\pm$ 0.17 & 1.3 & 0.7 \\
        1.00 $\leq z <$ 1.20 & 10.32 $\pm$ 0.39 & -2.14 $\pm$ 0.24 & 1.3 & 0.7 \\
        1.20 $\leq z <$ 1.70 & 10.88 $\pm$ 0.71 & -2.36 $\pm$ 0.05 & 1.3 & 0.7 \\
        1.70 $\leq z <$ 2.00 & 11.37 $\pm$ 0.19 & -2.61 $\pm$ 0.14 & 1.3 & 0.7 \\
        2.00 $\leq z <$ 2.50 & 11.21 $\pm$ 0.43 & -2.61 $\pm$ 0.08 & 1.3 & 0.7 \\
        2.50 $\leq z <$ 3.00 & 11.18 $\pm$ 0.54 & -2.56 $\pm$ 0.15 & 1.3 & 0.7 \\
        3.00 $\leq z <$ 4.20 & 11.57 $\pm$ 0.47 & -2.75 $\pm$ 0.17 & 1.3 & 0.7 \\
        4.20 $\leq z <$ 6.00 & 12.16 $\pm$ 0.05 & -3.60 $\pm$ 0.27 & 1.3 & 0.7 \\     
        \hline 
        CIGALE AGN \\
        \hline 
        0.00 $\leq z <$ 0.30 &  7.85 $\pm$ 1.08 & -2.34 $\pm$ 0.16 & 1.3 & 1.4 \\
        0.30 $\leq z <$ 0.45 &  8.10 $\pm$ 0.33 & -2.30 $\pm$ 0.12 & 1.3 & 1.3 \\
        0.45 $\leq z <$ 0.60 &  8.66 $\pm$ 0.20 & -2.32 $\pm$ 0.08 & 1.3 & 1.2 \\
        0.60 $\leq z <$ 0.80 &  8.84 $\pm$ 0.17 & -2.27 $\pm$ 0.08 & 1.3 & 1.1 \\
        0.80 $\leq z <$ 1.00 &  9.07 $\pm$ 0.24 & -2.23 $\pm$ 0.13 & 1.3 & 1.0 \\
        1.00 $\leq z <$ 1.20 &  9.34 $\pm$ 0.48 & -2.35 $\pm$ 0.06 & 1.3 & 0.9 \\
        1.20 $\leq z <$ 1.70 &  9.57 $\pm$ 0.44 & -2.18 $\pm$ 0.11 & 1.3 & 0.9 \\
        1.70 $\leq z <$ 2.00 &  9.82 $\pm$ 0.53 & -2.13 $\pm$ 0.11 & 1.3 & 0.9 \\
        2.00 $\leq z <$ 2.50 & 10.07 $\pm$ 0.62 & -2.21 $\pm$ 0.14 & 1.3 & 0.8 \\
        2.50 $\leq z <$ 3.00 & 10.46 $\pm$ 0.89 & -2.39 $\pm$ 0.09 & 1.3 & 0.7 \\
        3.00 $\leq z <$ 4.20 & 10.57 $\pm$ 0.35 & -2.51 $\pm$ 0.15 & 1.3 & 0.7 \\
        4.20 $\leq z <$ 6.00 & 10.82 $\pm$ 0.37 & -3.18 $\pm$ 0.29 & 1.3 & 0.7 \\
        \hline
        CIGALE SF \\
        \hline
        0.00 $\leq z <$ 0.30 &  8.66 $\pm$ 0.51 & -1.40 $\pm$ 0.07 & 1.3 & 1.1 \\
        0.30 $\leq z <$ 0.45 &  8.83 $\pm$ 0.17 & -1.43 $\pm$ 0.22 & 1.3 & 1.0 \\
        0.45 $\leq z <$ 0.60 &  9.19 $\pm$ 0.23 & -1.55 $\pm$ 0.18 & 1.3 & 0.9 \\
        0.60 $\leq z <$ 0.80 &  9.83 $\pm$ 0.30 & -1.66 $\pm$ 0.14 & 1.3 & 0.8 \\
        0.80 $\leq z <$ 1.00 & 10.05 $\pm$ 0.28 & -1.79 $\pm$ 0.24 & 1.3 & 0.7 \\
        1.00 $\leq z <$ 1.20 & 10.25 $\pm$ 0.29 & -1.99 $\pm$ 0.19 & 1.3 & 0.7 \\
        1.20 $\leq z <$ 1.70 & 11.08 $\pm$ 0.24 & -2.34 $\pm$ 0.18 & 1.3 & 0.7 \\
        1.70 $\leq z <$ 2.00 & 11.58 $\pm$ 0.27 & -2.53 $\pm$ 0.07 & 1.3 & 0.7 \\
        2.00 $\leq z <$ 2.50 & 11.63 $\pm$ 0.28 & -2.61 $\pm$ 0.02 & 1.3 & 0.7 \\
        2.50 $\leq z <$ 3.00 & 11.54 $\pm$ 0.25 & -2.70 $\pm$ 0.08 & 1.3 & 0.7 \\
        3.00 $\leq z <$ 4.20 & 11.40 $\pm$ 0.41 & -2.87 $\pm$ 0.02 & 1.3 & 0.7 \\
        4.20 $\leq z <$ 6.00 & 11.57 $\pm$ 0.03 & -3.43 $\pm$ 0.32 & 1.3 & 0.7      
        \botrule
    \end{tabular}
    \end{center}
\end{table}

Our ZFOURGE free parameters $L^{*}$ and $\phi^{*}$ exhibit evolutionary trends that differ from those reported in the literature. However, our fitting process is robust, and the evolution of the free parameters does not change significantly when the fixed parameters are altered. A known challenge in this analysis is the degeneracy between $L^{*}$ and $\phi^{*}$. A decrease in $L^{*}$ can be somewhat compensated for by increasing $\phi^{*}$ and vice-versa. Thus, the absolute values of the parameters themselves can be overlooked in favour of the overall trend in the dataset. 

As shown below, we find rapid evolution of ZFOURGE $L^{*}$ up to $z\approx2$, after which $L^{*}$ evolution slows. ZFOURGE $\phi^{*}$ shows consistent evolution in all redshift bins: 

\begin{equation*}
    L^{*}_{ZF} =
    \begin{cases} 
        10^{8.02 \pm 0.11} \times (1+z)^{7.36 \pm 0.25} & \text{for } z < 2, \\
        10^{8.65 \pm 0.31} \times (1+z)^{4.47 \pm 0.40} & \text{for } z > 2.
    \end{cases}
\end{equation*}

\begin{equation*}
    \phi^{*}_{ZF} =
    \begin{cases} 
        10^{-1.29 \pm 0.04} \times (1+z)^{-2.47 \pm 0.26} & \text{for } z < 2, \\
        10^{-1.57 \pm 0.59} \times (1+z)^{-1.91 \pm 1.04} & \text{for } z > 2.
    \end{cases}
\end{equation*}

Compared to the literature, both \cite{gruppioni_herschel_2013} and \cite{magnelli_deepest_2013} find much shallower $L^{*}$ evolution from $0<z<1$. ZFOURGE evolves 2-3x faster at $0<z<1$, but only 1.25x faster at $z>1$. This could be explained by ZFOURGE probing fainter luminosities. However, it is important to note that ZFOURGE was optimised for studying galaxies at $z>1$, where its deep near-infrared coverage is particularly effective. Consequently, results at $z<1$ should be interpreted cautiously, as the survey's design is less tailored to these lower redshifts. 

The CIGALE SF $L^{*}$ presents similar evolution compared to the ZFOURGE from $0<z<2$. In contrast to ZFOURGE results, the CIGALE SF $L^{*}$ declines from $z>2$ onwards. 

\begin{equation*}
    L^{*}_{SF} =
    \begin{cases} 
        10^{7.91 \pm 0.16} \times (1+z)^{8.07 \pm 0.37} & \text{for } z < 2, \\
        10^{11.55 \pm 0.47} \times (1+z)^{-0.06 \pm 0.73} & \text{for } z > 2.
    \end{cases}
\end{equation*}

This reversal is not seen in the evolution of $\phi^{*}$, which has a similar slope across all redshifts and is almost identical to ZFOURGE. 

\begin{equation*}
    \phi^{*}_{SF} =
    \begin{cases} 
        10^{-1.22 \pm 0.05} \times (1+z)^{-2.08 \pm 0.33} & \text{for } z < 2, \\
        10^{-1.48 \pm 0.46} \times (1+z)^{-2.21 \pm 0.81} & \text{for } z > 2.
    \end{cases}
\end{equation*}

The reversal in the evolution of $L^{*}$ below $z>2$ indicates that SF grew from at least $z=5$ to $z=2$ and has been declining ever since. It is well known (\citealp{gruppioni_herschel_2013,  madau_cosmic_2014} and references within) that the IR luminosity density has been declining since $z\approx2$ and we explore this more in Section \ref{Sec: IR Density}. 


We consistently see that, according to the evolution of $\phi^{*}$, the number of SF galaxies has increased since cosmic dawn. However, the characteristic luminosity $L^{*}$, has been declining since $z\approx2$. This shows that fainter SF galaxies become more common in the local universe. Our results reaffirm that $z\approx2$ is an important epoch for galaxy evolution.

The CIGALE AGN $L^{*}$ and $\phi^{*}$ evolve differently from the literature and from our ZFOURGE and CIGALE SF parameters at $z<2$:

\begin{equation*}
    L^{*}_{AGN} =
    \begin{cases} 
        10^{7.98 \pm 0.07} \times (1+z)^{4.06 \pm 0.17} & \text{for } z < 2, \\
        10^{9.15 \pm 0.24} \times (1+z)^{2.14 \pm 0.33} & \text{for } z > 2.
    \end{cases}
\end{equation*}

\begin{equation*}
    \phi^{*}_{AGN} =
    \begin{cases} 
        10^{-2.4 \pm 0.05} \times (1+z)^{0.52 \pm 0.16} & \text{for } z < 2, \\
        10^{-0.87 \pm 0.29} \times (1+z)^{-2.61 \pm 0.52} & \text{for } z > 2.
    \end{cases}
\end{equation*}

At $z>2$, the evolution of $L^{*}$ follows a similar trend to ZFOURGE, and a slower evolution at $z<2$. $\phi^{*}$ evolves similarly to SF and ZFOURGE at $z>2$. However, we see a reversal in the evolution of $\phi^{*}$ at $z<2$. This anomalous behaviour could be attributed to the degeneracy between $L^{*}$ and $\phi^{*}$. However, the magnitude of this degeneration has not been observed so prominently in the literature, suggesting two possibilities: CIGALE reveals a significant evolutionary epoch for AGN at $z<2$, or there is a bias in our fitting process. Our fitting process produces similar results for the ZFOURGE as seen in the literature and recovers the peak and turnover in the CIGALE SF, proving that $z<2$ is a significant epoch for AGN evolution. \cite{hopkins_observational_2007} and \cite{delvecchio_tracing_2014} show a similar reversal and subsequent decline in $\phi^{*}$ evolution below $z=1$. \cite{katsianis_evolution_2017} shows that high SFG rapidly decline below $z\approx1$ because of AGN feedback. We are therefore confident in our CIGALE AGN results. Our results imply fewer AGN in the local universe and are significantly fainter than in the early universe. 

Our results are complex, but showcase the importance of decomposing the SED of galaxies to separate the SF and AGN components. These trends indicate significant shifts in AGN and SF activity over cosmic time that were not detected in the combined ZFOURGE evolution.

\subsection{Bolometric IR Luminosity Density} \label{Sec: IR Density}
\begin{figure*}[t!]
    \centering
    \includegraphics[width=\textwidth]{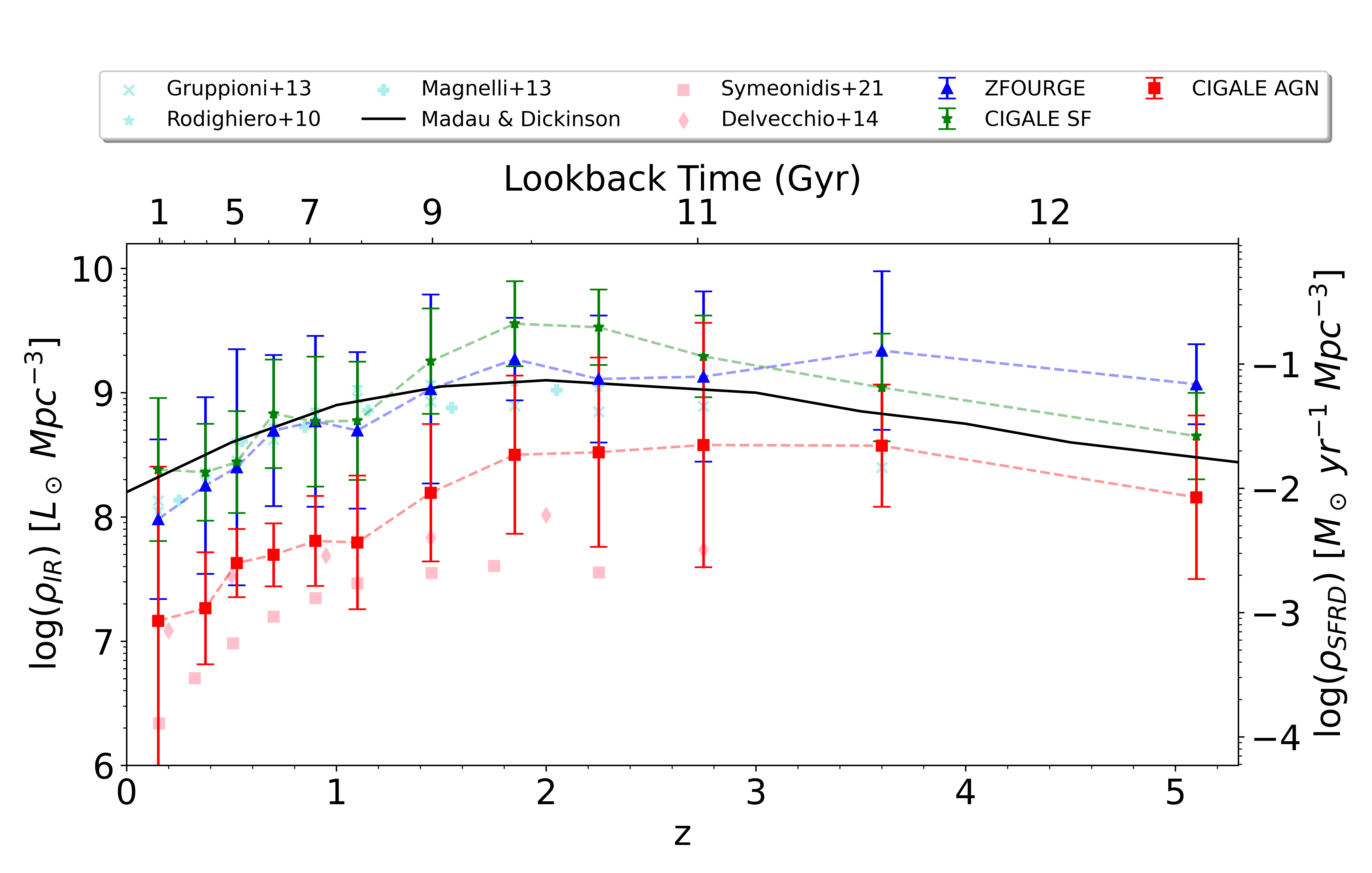}
    \caption{Evolution of the IR luminosity density (LD) calculated by integrating under the best fitting LFs. Uncertainties are calculated by re-performing the integration with errors from the LF fitting process. Blue triangles represent ZFOURGE; green stars CIGALE SF; and red squares CIGALE AGN. The right side y-axis is obtained from \cite{kennicutt_global_1998} based on a Salpeter IMF with $\rho_{SFRD} = \rho_{IR} \times 1.7\times10^{-10} \ L_{\odot}$. The top axis shows the lookback time in billions of years. We compare our results with relevant literature. \cite{gruppioni_herschel_2013, rodighiero_mid-_2010, magnelli_deepest_2013} as light blue compare the SF LD. \cite{symeonidis_agn_2021} and \cite{delvecchio_tracing_2014} as light red compare the AGN LD. The solid black line is the \cite{madau_cosmic_2014} LD.}
    \label{Fig: SFRD}
\end{figure*}

In this section, we calculate and analyse the bolometric IR (8-1000$\mu$m) luminosity density (LD) and star formation rate density ($\rho_{SFRD}$). Figure \ref{Fig: SFRD} shows the IR LD ($\rho_{IR}$) of our ZFOURGE and CIGALE results with values in table \ref{Tab: SFRD}. At each redshift bin, $\rho_{IR}$ is calculated by integrating under the best fitting luminosity function with equation \ref{EQ: Luminosity Density}:

\begin{equation} 
    \rho_{IR} = \frac{1}{\ln(10)} \int_{0}^{\infty} \varphi(L) dL 
    \label{EQ: Luminosity Density}
\end{equation}

Where $\phi(L)$ is the best fitting luminosity function (The Saunders function: Equation \ref{EQ: Saunders Function}). We utilise \texttt{scipy.integrate.quad} \citep{virtanen_scipy_2020} which uses an adaptive quadrature algorithm automatically subdividing the integration interval and applying a recursive Simpson’s rule. We perform the integration from $0$ to $\infty$ $L_{\odot}$ by cumulatively summing the integrand at incremental bounds (e.g. from $10^{10}$ to $10^{12}$ $L_{\odot}$, $10^{12}$ to $10^{14}$ $L_{\odot}$, etc) because the quadrature algorithm isn't well suited for small areas over very large bounds. In practice, additional calculations of $\rho_{IR}$ above $10^{20}$ $L_{\odot}$ are negligible. To generate $\rho_{IR}$ uncertainty values, we re-perform the integration using the LF fit errors.

\begin{figure*}
    \centering
    \includegraphics[width=\textwidth]{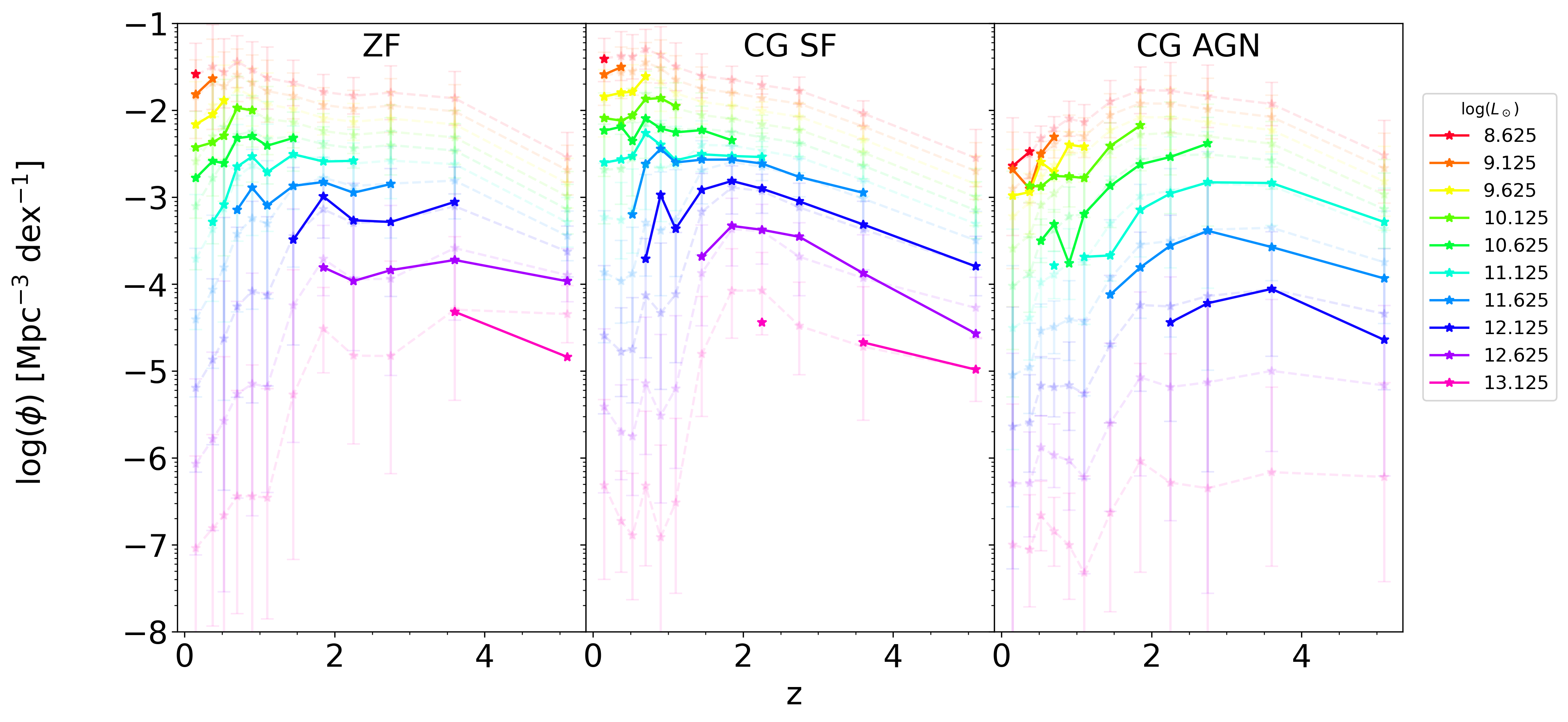}
    \caption{Luminosity class evolution as a function of redshift. $\phi$ values connected by straight lines correspond to real values in Figure \ref{Fig: Bolometric IR LF}. $\phi$ values connected by dashed lines are estimated from the best fitting LF. Error bars represent the propagated uncertainty derived from the LF. Real luminosity classes are 0.25 log$(L_{\odot})$ in width and centred in the middle (e.g. 8.5 --- 8.75 is centred on 8.625). Estimated classes are calculated at the centre of the luminosity bin (e.g. 8.625). Not all luminosity bins from Figure \ref{Fig: Bolometric IR LF} are displayed to reduce clutter.}
    \label{Fig: Class Evo}
\end{figure*}

The secondary right-hand-side axis of figure \ref{Fig: SFRD} displays the conversion to SFRD provided by \cite{kennicutt_global_1998} based on a Salpeter IMF calculated with $\rho_{SFRD} = \rho_{IR} \times 1.7\times10^{-10} L_{\odot}$. We remind the reader that the IR AGN densities do not have an associated SFR. The top x-axis shows the lookback time in billions of years, placing the evolution of the universe in the context of time to showcase the important evolutionary epochs.

Our ZFOURGE results in figure \ref{Fig: SFRD} show rapid evolution from $0<z<2$. From $z>2$ onwards, there is essentially no evolution as $\rho_{IR}$ remains roughly constant. As discussed in section \ref{Sec: Parameter Evolution}, our highest redshift bin was thought to suffer from incompleteness, but our $\rho_{IR}$ values are still high. We see excellent agreement with the literature from $0<z<2$ but deviate significantly from $z>2$ onwards. Importantly, we do not find a turnover in the IR LD at $z\approx2$ for ZFOURGE. This is a different result than is often published in the literature (\citealp{gruppioni_herschel_2013, magnelli_deepest_2013, madau_cosmic_2014, lutz_far-infrared_2014} and references within). However, this is not a new result as is seen in \cite{rodighiero_mid-_2010}, but they do not probe to a sufficiently high enough redshift to capture the decline above $z>2$.

The CIGALE decomposed SF IR LD is seen to increase from $0<z<2$ and decline from $z>2$ onwards. This agrees well with the literature, especially \cite{madau_cosmic_2014}. However, at $z>2$, our CIGALE results are elevated by $\approx 1.3$x. Some of this elevation can be attributed due to the higher bright-end slope $(\sigma = 0.7)$ compared to the literature $(\sigma = 0.5)$ (e.g. \citealp{gruppioni_herschel_2013}). However, this is not the sole reason. Likely due to poor FIR constraints, the luminosity is overestimated. As mentioned in Section \ref{Sec: ZF Total Discussion}, a drop in literature LFs is noticeable at $1.7\leq z< 2.0$. However, we do not find this drop in either the ZFOURGE or CIGALE datasets. Subsequently, $\rho_{IR}$ is elevated at $1.7\leq z< 2.0$ even when the bright end slope is reduced.  

The ability of CIGALE to recover the turnover in the SF $\rho_{IR}$, where ZFOURGE does not, suggests that CIGALE is effectively isolating the SF component and lends confidence to the accuracy of our results. The CIGALE SF LD is slightly elevated over ZFOURGE at most redshift bins. We believe this is not an error on CIGALE's part, but likely the FIR-poor ZFOURGE data it is based on. We find excellent agreement with the literature up until $z=2$. As mentioned previously in section \ref{Sec: Bolometric IR LF}, both \cite{fu_decomposing_2010} and \cite{wu_mid-infrared_2011} argue that when AGN are removed, the luminosity function is better fit with a Schechter function. Our CIGALE SF LF instead applies the Saunders function so that the CIGALE AGN and ZFOURGE LFs can be directly compared. This may also explain why our CIGALE SF LFs (figure \ref{Fig: Bolometric IR LF}) are elevated, as integrating under the Schechter functions results in a slight drop in $\rho_{IR}$. Furthermore, the uncertainties are almost all within the range of \cite{madau_cosmic_2014}.

The CIGALE AGN LD (or $\Psi_{BHAR}$) follows a similar evolution to the literature, though it is elevated, likely due to CIGALE's ability to detect fainter AGN. The trend seen in the CIGALE SF LD increasing from $0<z<2$ and declining from $z>2$ is still present, but not clear. This is not the first time a turnover in the AGN $\rho_{IR}$ has been seen. \cite{symeonidis_agn_2021} presents their IR AGN densities up to $z\approx2.5$. These results are at most $\approx 1$ order of magnitude lower than ours. We attribute this to their use of \cite{aird_evolution_2015} X-ray sources, which are converted to optical luminosity and then to IR luminosity. Their X-ray-selected galaxies likely miss the highly obscured and faint-luminosity counterpart that this work recovers. As \cite{symeonidis_agn_2021} only extends as far as $z\approx2.5$, the AGN $\rho_{IR}$ turnover is poorly defined. AGN by \cite{delvecchio_tracing_2014} agrees well with our results. We recalculated the AGN $\rho_{IR}$ for the \cite{delvecchio_tracing_2014} dataset because they did not provide $\rho_{IR}$ values in their work, instead focusing on the black hole accretion rate density ($\Psi_{bhar}$). Using the function parameters they reported, we use the same integration method described previously to calculate $\rho_{IR}$ and find good agreement with our results. Again, we attribute elevated densities to CIGALE's ability to recover fainter AGN.

Both \cite{symeonidis_agn_2021} and \cite{delvecchio_tracing_2014} show AGN LD peaks at $z\approx2$, but are poorly defined. Our CIGALE AGN LD peaks sometime between $z\approx2.5-3.5$. As was mentioned in section \ref{Sec: Parameter Evolution}, there appeared to be a significant evolutionary epoch for AGN occurring below sometime $z\approx2$ and this is reflected in our AGN LD results. From $0<z<2$, AGN density is seen to decline. Relying solely on the LD evolution of AGN and SF galaxies to assess their overall evolution may oversimplify their complex evolution. The functional fits have likely smoothed out slight variations in the LF, potentially masking essential details. 





\subsection{Space Density Evolution}
\label{Sec: Class Density}

In this section, we inspect the evolution of different luminosity classes by visualising the number density of each luminosity bin across redshift (figure \ref{Fig: Class Evo}). By incorporating the class density evolution into our analysis, a better picture of the evolution of galaxies and the co-evolution of AGN can be ascertained. When possible, the $\phi$ values are taken from the existing luminosity bins\footnote{$\phi$ values ending in X.875 and X.375 $\log_{10}(L_{\odot})$ are skipped. Shown bins are 0.25 $\log_{10}(L_{\odot})$ wide.}. Otherwise, $\phi$ values are calculated from the best-fitting LF. The class density evolution (figure \ref{Fig: Class Evo}) can be thought of as the transpose of the LF (figures \ref{Fig: Bolometric IR LF} and \ref{Fig: LF Filled}). The LF and class density are complementary because they allow us to view the number density as an evolution with luminosity and redshift, respectively. 

\begin{table}[h]
    \begin{center}
    \caption{Luminosity density as a function of redshift. Units are in log($\rho_{IR}$) [$L_{\odot}$ Mpc$^{-3}$]. $\rho_{IR}$ values are centered on the redshift bin.}
    \label{Tab: SFRD}
    \begin{tabular}{@{}cccc@{}}
        \toprule
        $z$ & ZFOURGE & CIGALE SF & CIGALE AGN \\
        \hline
        0.30 $\leq z <$ 0.45 & 7.98 $\pm$ 0.81 & 8.34 $\pm$ 0.66 & 7.16 $\pm$ 0.85 \\
        0.45 $\leq z <$ 0.60 & 8.25 $\pm$ 0.70 & 8.36 $\pm$ 0.40 & 7.24 $\pm$ 0.44 \\
        0.00 $\leq z <$ 0.30 & 8.40 $\pm$ 0.95 & 8.44 $\pm$ 0.42 & 7.63 $\pm$ 0.27 \\
        0.60 $\leq z <$ 0.80 & 8.69 $\pm$ 0.59 & 8.83 $\pm$ 0.44 & 7.69 $\pm$ 0.26 \\
        0.80 $\leq z <$ 1.00 & 8.77 $\pm$ 0.73 & 8.77 $\pm$ 0.56 & 7.81 $\pm$ 0.36 \\
        1.00 $\leq z <$ 1.20 & 8.70 $\pm$ 0.68 & 8.77 $\pm$ 0.49 & 7.79 $\pm$ 0.55 \\
        1.20 $\leq z <$ 1.70 & 9.03 $\pm$ 0.80 & 9.25 $\pm$ 0.44 & 8.19 $\pm$ 0.58 \\
        1.70 $\leq z <$ 2.00 & 9.27 $\pm$ 0.34 & 9.56 $\pm$ 0.37 & 8.50 $\pm$ 0.68 \\
        2.00 $\leq z <$ 2.50 & 9.11 $\pm$ 0.54 & 9.53 $\pm$ 0.32 & 8.52 $\pm$ 0.79 \\
        2.50 $\leq z <$ 3.00 & 9.13 $\pm$ 0.67 & 9.35 $\pm$ 0.39 & 8.58 $\pm$ 1.03 \\
        3.00 $\leq z <$ 4.20 & 9.34 $\pm$ 0.65 & 9.04 $\pm$ 0.48 & 8.57 $\pm$ 0.51 \\
        4.20 $\leq z <$ 6.00 & 9.07 $\pm$ 0.34 & 8.65 $\pm$ 0.36 & 8.16 $\pm$ 0.66 
        \botrule
    \end{tabular}
    \end{center}        
\end{table}

In figure \ref{Fig: Class Evo}, we present IR luminosity classes as low as $L_{IR}=10^{8.5}\ L_{\odot}$. We find that the space density of ZFOURGE LIRGs and ULIRGs have been consistently declining since at least $z=2$, and likely even earlier for ULIRGs. Galaxies fainter than LIRGs (FIRGs, $L_{IR} < 10^{11} L_{\odot}$) evolve differently, beginning to decline at a lower redshift than their brighter luminosity counterparts. The redshift at which galaxies begin declining in number density is related to their luminosity. ZFOURGE galaxies fainter than $L_{IR} < 10^{9}\ L_{\odot}$ appear to be increasing in number density across all of cosmic time and have yet to begin declining. We find similar agreement in the literature with \cite{rodighiero_mid-_2010} and \cite{gruppioni_herschel_2013} with our results mostly in agreement. We attribute the differences to slight variations in the classes and methods within. FIRGs dominate the ZFOURGE LD from $0<z<1.2$, declining from 74\% to 47\%. LIRGs dominate from $1.2<z<3$, remaining steady with 43\% to 51\% contribution. At $z>3$, ULIRGs dominate LD, increasing rapidly from 38\% to 74\%. In the highest redshift bin, FIRG contribution drops to 5\%

CIGALE SF galaxies evolve differently from ZFOURGE, although similar contributions to the LD are seen. FIRGs dominate LD density from $0<z<1.0$, remaining roughly constant between 49\% to 63\%. LIRGs only dominate LD from $1.0<z<1.2$ with 45\% contribution. ULIRGs dominate LD from $z>1.2$ onwards, with contribution increasing from 48\% to 75\%. Figure \ref{Fig: Class Evo} shows that SF LIRGs evolve similarly to FIRGs at $z>2$. However, the estimated $\phi$ values show an increasing number density with decreasing luminosity for all FIRGs. A possible evolution scenario is theorised for SFG: all evolve similarly from high redshift to $z \approx 2$, increasing in number density from high redshift. At and below $z \approx 2$, the brightest FIRG number densities begin to peak and decline earlier than fainter FIRG counterparts, which have yet to start declining. LIRGs decline faster and earlier than FIRGs, and ULIRGs decline faster and sooner than LIRGs. This reflects a \textit{downsizing} scenario in which brighter galaxies peak in number density at higher redshift \citep{merloni_synthesis_2008, wylezalek_galaxy_2014, fiore_agn_2017}.

CIGALE AGN again evolve differently. Faint IR AGN dominate LD from $0 \leq z < 1.7$ and luminous IR AGN dominate LD from $z>1.2$ onwards. Ultra-luminous IR AGN never dominates. The difference between FIRGs and LIRGs is stark. There is a clear, systematic shift in the peak number density with luminosity class. The \textit{downsizing} seen in SF galaxies is more pronounced in AGN. As AGN luminosity increases, the number density of AGN peaks at higher redshifts and declines earlier than their fainter luminosity counterparts. When comparing the \textit{downsizing} effect between SFGs and AGN, it is unmistakable that galaxies with a luminous AGN decline faster and earlier than equally bright SFG counterparts.

\section{Conclusion}

We utilised ZFOURGE data as inputs for CIGALE to separate the SF and AGN components of the infrared spectral energy distribution to create and analyse their respective luminosity functions. Our results are summarised as follows:

\begin{enumerate}[(i)]
    \item We first generate LFs of ZFOURGE galaxies from a sub-sample of 18,373 galaxies that accounted for incompleteness and bias across various redshift bins ranging from $0<z<6$.
    
    \item After CIGALE decomposition, we find 21,390 sources (28.9\% of the sample) with a significant AGN fraction ($\mathcal{F}_{AGN} > 0.1$). This was reduced to 16,850 sources (22.8\%) after an 80\% flux completion cut. In context with AGN dominant sources removed in ZFOURGE by \cite{cowley_zfourge_2016}, simple MIR colour-colour selection diagnostics are insufficient to identify low luminosity AGN, which we show significantly influences galaxy samples.
    
    \item We use a Saunders function to fit CIGALE SF LFs, although a Schechter function may fit too, as predicted by \cite{wu_mid-infrared_2011, fu_decomposing_2010}. Brighter SF galaxies at higher redshifts are required to confirm. 

    \item A peak and turnover in the evolution of CIGALE SF $L^{*}$ is found at $z \approx 2$ coinciding with the peak of cosmic star formation. Similarly, SF LD is also seen to peak and decline above $z\approx2$. This agrees with established literature and provides good evidence supporting our CIGALE results. Importantly, no turnover in ZFOURGE LD was found, highlighting the importance of SED decomposition.
    
    \item A peak and turn over in $\phi^{*}$ is seen for CIGALE AGN at $z \approx 2$ with $L^{*}$ continuing to decline. Although the AGN LD declines from $z<2$, above $z>2$ does not show a defined peak. The peak of AGN LF likely lies between $z\approx2.5-3.5$. \cite{delvecchio_tracing_2014} finds a similar trend at the same time for their AGN $\phi^{*}$, but declines below $z<1$. \cite{katsianis_evolution_2017} also shows AGN feedback significantly affects SF at $z<1$. We have found reliable evidence that $z\approx2$ is a significant evolutionary epoch for AGN evolution. 
    
    \item In context, $L^{*}_{SF}$ peaks at the same time as $\phi^{*}_{AGN}$ while $\phi^{*}_{SF}$ increases across all cosmic time. These results show fainter SF galaxies becoming increasingly common in the local universe while brighter SF galaxies trend towards extinction.  

    \item The space density evolution of luminosity classes for SFG and AGN are very different. SF LIRGs and ULIRGs increase in number density and peak at $z \approx 2$ before rapidly declining to the present day. SF FIRGs evolve similarly across fainter luminosities and peak at a lower redshift. The faintest SF FIRGs have yet to peak in number density. AGN number densities peak much earlier in the universe than SF counterparts. Luminous and ultraluminous AGN have declined since at least $z \approx 3$ and likely even earlier. There is a strong correlation between peak AGN density and luminosity. Fainter AGN peak much later in the universe and have declined since $z \approx 2$. This shows a clear \textit{downsizing} effect which is more prominent in AGN.

    \item From high redshift until $z\approx2$, SFGs form with increasing brightness and consume a significant fraction of the available gas supply. Below $z<2$, the faintest SFG grow in number, but not as big and bright at $z<2$ as the gas supply becomes increasingly exhausted. This is reflected in our class evolution results and is the likely reason behind the \textit{downsizing} effect.

    \item The brightest AGN are already in decline at the beginning of the universe. These AGN decline at a consistent rate until $z\approx2$. Conversely, the faintest AGN increase in number density until $z\approx2$. Below $z=2$, all AGN show a rapid decline. At $z\approx1$, the brightest AGN show signs of increasing in number. This increase appears to last until $z\approx0.5$ before declining again. This reflects a period where the available gas supply shifted to fueling AGN growth.
\end{enumerate}

These results indicate a scenario where the available gas supply favoured SFG from high redshift until $z\approx2$ when the supply becomes increasingly exhausted. Below $z<2$, the remaining gas predominantly formed fainter SFG, with bright SFG rapidly declining to near extinction by $z\approx1$ when number densities stabilise. By $z=1$, the brightest AGN are almost as equally common as the brightest SFG. Given that AGN of a particular brightness decline sooner than equally bright SFG counterparts, it is probable that AGN positively influences the growth of SF.

\section*{Acknowledgements}
This research was supported by an Australian Government Research Training Program (RTP) Scholarship. Thank you to the anonymous referee, who provided valuable comments that improved this research. Thank you to our colleague Vanessa Porchet for using CIGALE to run a reduced-band analysis. Thank you to Vera Delfavero for helpful comments.

\section*{Data Availability}
Python notebooks and scripts that analysed the data are available on GitHub at \url{https://github.com/daniel-lyon/MPhil-Code}

\bibliography{references.bib}
\bibliographystyle{aasjournal}

\onecolumn

\section*{Appendix A} \label{Sec: Appendix}

\begin{figure}[h!]
    \includegraphics[width=0.95\textwidth]{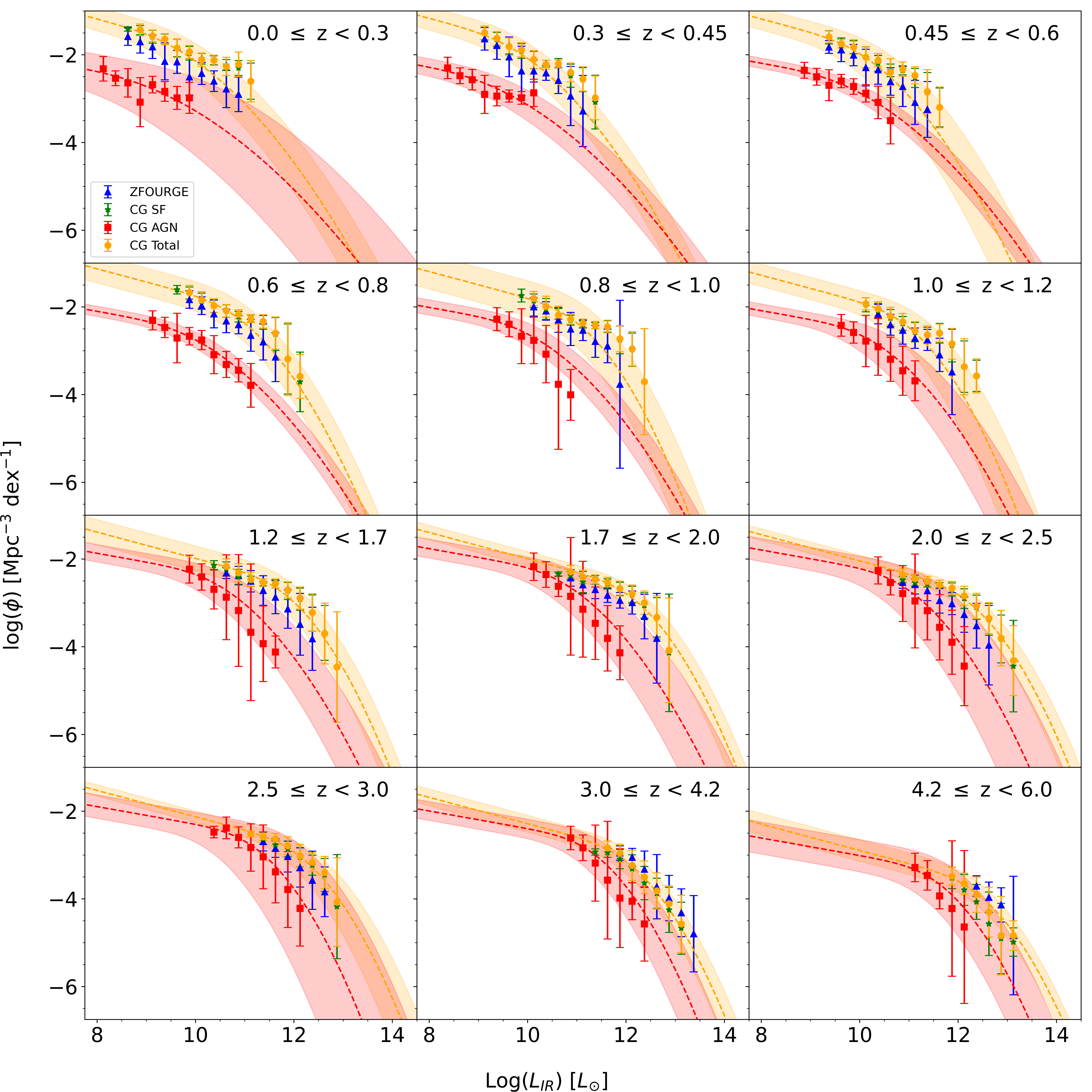}
    \caption{The bolometric IR $(8-1000\mu m)$ LF of ZFOURGE (Blue), CIGALE SF (Green), CIGALE AGN (Red), CIGALE Total (Orange).}
    \label{Fig: Appendix LF}
\end{figure}

\end{document}